
%
%
%
\def\unredoffs{} \def\redoffs{\voffset=-.31truein\hoffset=-.48truein}
\def\speclscape{}
%
%
%
%
%
\newbox\leftpage \newdimen\fullhsize \newdimen\hstitle \newdimen\hsbody
\tolerance=1000\hfuzz=2pt
\catcode`\@=11 
\ifx\hyperdef\UNd@FiNeD\def\hyperdef#1#2#3#4{#4}\def\hyperref#1#2#3#4{#4}\fi
\def\bigans{b }
\def\answ{b }
%
\ifx\answ\bigans\message{(This will come out unreduced.}
\magnification=1200\unredoffs\baselineskip=16pt plus 2pt minus 1pt
\hsbody=\hsize \hstitle=\hsize 
\else\message{(This will be reduced.} \let\l@r=L
\magnification=1000\baselineskip=16pt plus 2pt minus 1pt \vsize=7truein
\redoffs \hstitle=8truein\hsbody=4.75truein\fullhsize=10truein\hsize=\hsbody
\output={\ifnum\pageno=0 
  \shipout\vbox{\speclscape{\hsize\fullhsize\makeheadline}
    \hbox to \fullhsize{\hfill\pagebody\hfill}}\advancepageno
  \else
  \almostshipout{\leftline{\vbox{\pagebody\makefootline}}}\advancepageno
  \fi}
\def\almostshipout#1{\if L\l@r \count1=1 \message{[\the\count0.\the\count1]}
      \global\setbox\leftpage=#1 \global\let\l@r=R
 \else \count1=2
  \shipout\vbox{\speclscape{\hsize\fullhsize\makeheadline}
      \hbox to\fullhsize{\box\leftpage\hfil#1}}  \global\let\l@r=L\fi}
\fi
%
\newcount\yearltd\yearltd=\year\advance\yearltd by -2000

\def\Title#1#2{\nopagenumbers\abstractfont\hsize=\hstitle\rightline{#1}%
\vskip 1in\centerline{\titlefont #2}\abstractfont\vskip .5in\pageno=0}
\def\Date#1{\vfill\leftline{#1}\tenpoint\supereject\global\hsize=\hsbody%
\footline={\hss\tenrm\hyperdef\hypernoname{page}\folio\folio\hss}}%
%

\def\draftmode{\message{ DRAFTMODE }\def\draftdate{{\rm preliminary draft:
\number\month/\number\day/\number\yearltd\ \ \hourmin}}%
\headline={\hfil\draftdate}\writelabels\baselineskip=20pt plus 2pt minus 2pt
 {\count255=\time\divide\count255 by 60 \xdef\hourmin{\number\count255}
  \multiply\count255 by-60\advance\count255 by\time
  \xdef\hourmin{\hourmin:\ifnum\count255<10 0\fi\the\count255}}}
\def\nolabels{\def\wrlabeL##1{}\def\eqlabeL##1{}\def\reflabeL##1{}}
\def\writelabels{\def\wrlabeL##1{\leavevmode\vadjust{\rlap{\smash%
{\line{{\escapechar=` \hfill\rlap{\sevenrm\hskip.03in\string##1}}}}}}}%
\def\eqlabeL##1{{\escapechar-1\rlap{\sevenrm\hskip.05in\string##1}}}%
\def\reflabeL##1{\noexpand\llap{\noexpand\sevenrm\string\string\string##1}}}
\nolabels
%
\global\newcount\secno \global\secno=0
\global\newcount\meqno \global\meqno=1
\def\s@csym{}
\def\newsec#1{\global\advance\secno by1%
{\toks0{#1}\message{(\the\secno. \the\toks0)}}%
\global\subsecno=0\eqnres@t\let\s@csym\secsym\xdef\secn@m{\the\secno}\noindent
{\bf\hyperdef\hypernoname{section}{\the\secno}{\the\secno.} #1}%
\writetoca{{\string\hyperref{}{section}{\the\secno}{\the\secno.}} {#1}}%
\par\nobreak\medskip\nobreak}
\def\eqnres@t{\xdef\secsym{\the\secno.}\global\meqno=1\bigbreak\bigskip}
\def\sequentialequations{\def\eqnres@t{\bigbreak}}\xdef\secsym{}
\global\newcount\subsecno \global\subsecno=0
\def\subsec#1{\global\advance\subsecno by1%
{\toks0{#1}\message{(\s@csym\the\subsecno. \the\toks0)}}%
\ifnum\lastpenalty>9000\else\bigbreak\fi
\noindent{\it\hyperdef\hypernoname{subsection}{\secn@m.\the\subsecno}%
{\secn@m.\the\subsecno.} #1}\writetoca{\string\quad
{\string\hyperref{}{subsection}{\secn@m.\the\subsecno}{\secn@m.\the\subsecno.}}
{#1}}\par\nobreak\medskip\nobreak}
\def\appendix#1#2{\global\meqno=1\global\subsecno=0\xdef\secsym{\hbox{#1.}}%
\bigbreak\bigskip\noindent{\bf Appendix \hyperdef\hypernoname{appendix}{#1}%
{#1.} #2}{\toks0{(#1. #2)}\message{\the\toks0}}%
\xdef\s@csym{#1.}\xdef\secn@m{#1}%
\writetoca{\string\hyperref{}{appendix}{#1}{Appendix {#1.}} {#2}}%
\par\nobreak\medskip\nobreak}
%
%
\def\checkm@de#1#2{\ifmmode{\def\f@rst##1{##1}\hyperdef\hypernoname{equation}%
{#1}{#2}}\else\hyperref{}{equation}{#1}{#2}\fi}
\def\eqnn#1{\DefWarn#1\xdef #1{(\noexpand\relax\noexpand\checkm@de%
{\s@csym\the\meqno}{\secsym\the\meqno})}%
\wrlabeL#1\writedef{#1\leftbracket#1}\global\advance\meqno by1}
\def\f@rst#1{\c@t#1a\em@ark}\def\c@t#1#2\em@ark{#1}
\def\eqna#1{\DefWarn#1\wrlabeL{#1$\{\}$}%
\xdef #1##1{(\noexpand\relax\noexpand\checkm@de%
{\s@csym\the\meqno\noexpand\f@rst{##1}}{\hbox{$\secsym\the\meqno##1$}})}
\writedef{#1\numbersign1\leftbracket#1{\numbersign1}}\global\advance\meqno by1}
\def\eqn#1#2{\DefWarn#1%
\xdef #1{(\noexpand\hyperref{}{equation}{\s@csym\the\meqno}%
{\secsym\the\meqno})}$$#2\eqno(\hyperdef\hypernoname{equation}%
{\s@csym\the\meqno}{\secsym\the\meqno})\eqlabeL#1$$%
\writedef{#1\leftbracket#1}\global\advance\meqno by1}
\def\xeqn{\expandafter\xe@n}\def\xe@n(#1){#1}
\def\xeqna#1{\expandafter\xe@n#1}
\def\eqns#1{(\e@ns #1{\hbox{}})}
\def\e@ns#1{\ifx\UNd@FiNeD#1\message{eqnlabel \string#1 is undefined.}%
\xdef#1{(?.?)}\fi{\let\hyperref=\relax\xdef\next{#1}}%
\ifx\next\em@rk\def\next{}\else%
\ifx\next#1\xeqn#1\else\def\n@xt{#1}\ifx\n@xt\next#1\else\xeqna#1\fi
\fi\let\next=\e@ns\fi\next}

\def\DefWarn#1{\ifx\UNd@FiNeD#1\else
\immediate\write16{*** WARNING: the label \string#1 is already defined ***}\fi}
%
\newskip\footskip\footskip14pt plus 1pt minus 1pt 
\def\footnotefont{\ninepoint}\def\f@t#1{\footnotefont #1\@foot}
\def\f@@t{\baselineskip\footskip\bgroup\footnotefont\aftergroup\@foot\let\next}
\setbox\strutbox=\hbox{\vrule height9.5pt depth4.5pt width0pt}
\global\newcount\ftno \global\ftno=0
\def\foot{\global\advance\ftno by1\def\foot@rg{\hyperref{}{footnote}%
{\the\ftno}{\the\ftno}\xdef\foot@rg{\noexpand\hyperdef\noexpand\hypernoname%
{footnote}{\the\ftno}{\the\ftno}}}\footnote{$^{\foot@rg}$}}
%
\newwrite\ftfile
\def\footend{\def\foot{\global\advance\ftno by1\chardef\wfile=\ftfile
\hyperref{}{footnote}{\the\ftno}{$^{\the\ftno}$}%
\ifnum\ftno=1\immediate\openout\ftfile=\jobname.fts\fi%
\immediate\write\ftfile{\noexpand\smallskip%
\noexpand\item{\noexpand\hyperdef\noexpand\hypernoname{footnote}
{\the\ftno}{f\the\ftno}:\ }\pctsign}\findarg}%
\def\footatend{\vfill\eject\immediate\closeout\ftfile{\parindent=20pt
\centerline{\bf Footnotes}\nobreak\bigskip\input \jobname.fts }}}
\def\footatend{}
%
%
\global\newcount\refno \global\refno=1
\newwrite\rfile
\def\ref{[\hyperref{}{reference}{\the\refno}{\the\refno}]\nref}
\def\nref#1{\DefWarn#1%
\xdef#1{[\noexpand\hyperref{}{reference}{\the\refno}{\the\refno}]}%
\writedef{#1\leftbracket#1}%
\ifnum\refno=1\immediate\openout\rfile=\jobname.refs\fi
\chardef\wfile=\rfile\immediate\write\rfile{\noexpand\item{[\noexpand\hyperdef%
\noexpand\hypernoname{reference}{\the\refno}{\the\refno}]\ }%
\reflabeL{#1\hskip.31in}\pctsign}\global\advance\refno by1\findarg}
\def\findarg#1#{\begingroup\obeylines\newlinechar=`\^^M\pass@rg}
{\obeylines\gdef\pass@rg#1{\writ@line\relax #1^^M\hbox{}^^M}%
\gdef\writ@line#1^^M{\expandafter\toks0\expandafter{\striprel@x #1}%
\edef\next{\the\toks0}\ifx\next\em@rk\let\next=\endgroup\else\ifx\next\empty%
\else\immediate\write\wfile{\the\toks0}\fi\let\next=\writ@line\fi\next\relax}}
\def\striprel@x#1{} \def\em@rk{\hbox{}}
\def\lref{\begingroup\obeylines\lr@f}
\def\lr@f#1#2{\DefWarn#1\gdef#1{\let#1=\UNd@FiNeD\ref#1{#2}}\endgroup\unskip}

\def\addref#1{\immediate\write\rfile{\noexpand\item{}#1}} 
\def\listrefs{\footatend\vfill\supereject\immediate\closeout\rfile\writestoppt
\baselineskip=\footskip\centerline{{\bf References}}\bigskip{\parindent=20pt%
\frenchspacing\escapechar=` \input \jobname.refs\vfill\eject}\nonfrenchspacing}
\def\startrefs#1{\immediate\openout\rfile=\jobname.refs\refno=#1}
\def\xref{\expandafter\xr@f}\def\xr@f[#1]{#1}
\def\refs#1{\count255=1[\r@fs #1{\hbox{}}]}
\def\r@fs#1{\ifx\UNd@FiNeD#1\message{reflabel \string#1 is undefined.}%
\nref#1{need to supply reference \string#1.}\fi%
\vphantom{\hphantom{#1}}{\let\hyperref=\relax\xdef\next{#1}}%
\ifx\next\em@rk\def\next{}%
\else\ifx\next#1\ifodd\count255\relax\xref#1\count255=0\fi%
\else#1\count255=1\fi\let\next=\r@fs\fi\next}
%

%
\newwrite\ffile\global\newcount\figno \global\figno=1
\def\fig{fig.~\hyperref{}{figure}{\the\figno}{\the\figno}\nfig}
\def\nfig#1{\DefWarn#1%
\xdef#1{fig.~\noexpand\hyperref{}{figure}{\the\figno}{\the\figno}}%
\writedef{#1\leftbracket fig.\noexpand~\xfig#1}%
\ifnum\figno=1\immediate\openout\ffile=\jobname.figs\fi\chardef\wfile=\ffile%
{\let\hyperref=\relax
\immediate\write\ffile{\noexpand\medskip\noexpand\item{Fig.\ %
\noexpand\hyperdef\noexpand\hypernoname{figure}{\the\figno}{\the\figno}. }
\reflabeL{#1\hskip.55in}\pctsign}}\global\advance\figno by1\findarg}
\def\listfigs{\vfill\eject\immediate\closeout\ffile{\parindent40pt
\baselineskip14pt\centerline{{\bf Figure Captions}}\nobreak\medskip
\escapechar=` \input \jobname.figs\vfill\eject}}
\def\xfig{\expandafter\xf@g}\def\xf@g fig.\penalty\@M\ {}
\def\figs#1{figs.~\f@gs #1{\hbox{}}}
\def\f@gs#1{{\let\hyperref=\relax\xdef\next{#1}}\ifx\next\em@rk\def\next{}\else
\ifx\next#1\xfig #1\else#1\fi\let\next=\f@gs\fi\next}
\def\figin{\epsfcheck\figin}\def\figins{\epsfcheck\figins}
\def\epsfcheck{\ifx\epsfbox\UNd@FiNeD
\message{(NO epsf.tex, FIGURES WILL BE IGNORED)}
\gdef\figin##1{\vskip2in}\gdef\figins##1{\hskip.5in}
\else\message{(FIGURES WILL BE INCLUDED)}%
\gdef\figin##1{##1}\gdef\figins##1{##1}\fi}
\def\DefWarn#1{}
\def\figinsert{\goodbreak\midinsert}
\def\ifig#1#2#3{\DefWarn#1\xdef#1{fig.~\noexpand\hyperref{}{figure}%
{\the\figno}{\the\figno}}\writedef{#1\leftbracket fig.\noexpand~\xfig#1}%
\figinsert\figin{\centerline{#3}}\medskip\centerline{\vbox{\baselineskip12pt
\advance\hsize by -1truein\noindent\wrlabeL{#1=#1}\footnotefont%
{\bf Fig.~\hyperdef\hypernoname{figure}{\the\figno}{\the\figno}:} #2}}
\bigskip\endinsert\global\advance\figno by1}
\newwrite\lfile
{\escapechar-1\xdef\pctsign{\string\%}\xdef\leftbracket{\string\{}
\xdef\rightbracket{\string\}}\xdef\numbersign{\string\#}}
\def\writedefs{\immediate\openout\lfile=\jobname.defs \def\writedef##1{%
{\let\hyperref=\relax\let\hyperdef=\relax\let\hypernoname=\relax
 \immediate\write\lfile{\string\def\string##1\rightbracket}}}}%
\def\writestop{\def\writestoppt{\immediate\write\lfile{\string\pageno
 \the\pageno\string\startrefs\leftbracket\the\refno\rightbracket
 \string\def\string\secsym\leftbracket\secsym\rightbracket
 \string\secno\the\secno\string\meqno\the\meqno}\immediate\closeout\lfile}}
\def\writestoppt{}\def\writedef#1{}
\def\seclab#1{\DefWarn#1%
\xdef #1{\noexpand\hyperref{}{section}{\the\secno}{\the\secno}}%
\writedef{#1\leftbracket#1}\wrlabeL{#1=#1}}
\def\subseclab#1{\DefWarn#1%
\xdef #1{\noexpand\hyperref{}{subsection}{\secn@m.\the\subsecno}%
{\secn@m.\the\subsecno}}\writedef{#1\leftbracket#1}\wrlabeL{#1=#1}}
\def\applab#1{\DefWarn#1%
\xdef #1{\noexpand\hyperref{}{appendix}{\secn@m}{\secn@m}}%
\writedef{#1\leftbracket#1}\wrlabeL{#1=#1}}
\newwrite\tfile \def\writetoca#1{}
\def\leaderfill{\leaders\hbox to 1em{\hss.\hss}\hfill}
\def\writetoc{\immediate\openout\tfile=\jobname.toc
   \def\writetoca##1{{\edef\next{\write\tfile{\noindent ##1
   \string\leaderfill {\string\hyperref{}{page}{\noexpand\number\pageno}%
                       {\noexpand\number\pageno}} \par}}\next}}}
\newread\ch@ckfile
\def\listtoc{\immediate\closeout\tfile\immediate\openin\ch@ckfile=\jobname.toc
\ifeof\ch@ckfile\message{no file \jobname.toc, no table of contents this pass}%
\else\closein\ch@ckfile\centerline{\bf Contents}\nobreak\medskip%
{\baselineskip=12pt\footnotefont\parskip=0pt\catcode`\@=11\input\jobname.toc
\catcode`\@=12\bigbreak\bigskip}\fi}
\catcode`\@=12 
%
\edef\tfontsize{\ifx\answ\bigans scaled\magstep3\else scaled\magstep4\fi}
\font\titlerm=cmr10 \tfontsize \font\titlerms=cmr7 \tfontsize
\font\titlermss=cmr5 \tfontsize \font\titlei=cmmi10 \tfontsize
\font\titleis=cmmi7 \tfontsize \font\titleiss=cmmi5 \tfontsize
\font\titlesy=cmsy10 \tfontsize \font\titlesys=cmsy7 \tfontsize
\font\titlesyss=cmsy5 \tfontsize \font\titleit=cmti10 \tfontsize
\skewchar\titlei='177 \skewchar\titleis='177 \skewchar\titleiss='177
\skewchar\titlesy='60 \skewchar\titlesys='60 \skewchar\titlesyss='60
\def\titlefont{\def\rm{\fam0\titlerm}
\textfont0=\titlerm \scriptfont0=\titlerms \scriptscriptfont0=\titlermss
\textfont1=\titlei \scriptfont1=\titleis \scriptscriptfont1=\titleiss
\textfont2=\titlesy \scriptfont2=\titlesys \scriptscriptfont2=\titlesyss
\textfont\itfam=\titleit \def\it{\fam\itfam\titleit}\rm}
 \ifx\answ\bigans\else scaled\magstep1\fi
\ifx\answ\bigans\def\abstractfont{\tenpoint}\else
\font\absit=cmti10 scaled \magstep1
\font\abssl=cmsl10 scaled \magstep1
\font\absrm=cmr10 scaled\magstep1 \font\absrms=cmr7 scaled\magstep1
\font\absrmss=cmr5 scaled\magstep1 \font\absi=cmmi10 scaled\magstep1
\font\absis=cmmi7 scaled\magstep1 \font\absiss=cmmi5 scaled\magstep1
\font\abssy=cmsy10 scaled\magstep1 \font\abssys=cmsy7 scaled\magstep1
\font\abssyss=cmsy5 scaled\magstep1 \font\absbf=cmbx10 scaled\magstep1
\skewchar\absi='177 \skewchar\absis='177 \skewchar\absiss='177
\skewchar\abssy='60 \skewchar\abssys='60 \skewchar\abssyss='60
\def\abstractfont{\def\rm{\fam0\absrm}
\textfont0=\absrm \scriptfont0=\absrms \scriptscriptfont0=\absrmss
\textfont1=\absi \scriptfont1=\absis \scriptscriptfont1=\absiss
\textfont2=\abssy \scriptfont2=\abssys \scriptscriptfont2=\abssyss
\textfont\itfam=\absit \def\it{\fam\itfam\absit}\def\footnotefont{\tenpoint}%
\textfont\slfam=\abssl \def\sl{\fam\slfam\abssl}%
\textfont\bffam=\absbf \def\bf
{\fam\bffam\absbf}\rm}\fi
\def\tenpoint{\def\rm{\fam0\tenrm}
\textfont0=\tenrm \scriptfont0=\sevenrm \scriptscriptfont0=\fiverm
\textfont1=\teni  \scriptfont1=\seveni  \scriptscriptfont1=\fivei
\textfont2=\tensy \scriptfont2=\sevensy \scriptscriptfont2=\fivesy
\textfont\itfam=\tenit \def\it{\fam\itfam\tenit}\def\footnotefont{\ninepoint}%
\textfont\bffam=\tenbf \def\bf{\fam\bffam\tenbf}\def\sl{\fam\slfam\tensl}\rm}
\font\ninerm=cmr9 \font\sixrm=cmr6 \font\ninei=cmmi9 \font\sixi=cmmi6
\font\ninesy=cmsy9 \font\sixsy=cmsy6 \font\ninebf=cmbx9
\font\nineit=cmti9 \font\ninesl=cmsl9 \skewchar\ninei='177
\skewchar\sixi='177 \skewchar\ninesy='60 \skewchar\sixsy='60
\def\ninepoint{\def\rm{\fam0\ninerm}
\textfont0=\ninerm \scriptfont0=\sixrm \scriptscriptfont0=\fiverm
\textfont1=\ninei \scriptfont1=\sixi \scriptscriptfont1=\fivei
\textfont2=\ninesy \scriptfont2=\sixsy \scriptscriptfont2=\fivesy
\textfont\itfam=\ninei \def\it{\fam\itfam\nineit}\def\sl{\fam\slfam\ninesl}%
\textfont\bffam=\ninebf \def\bf{\fam\bffam\ninebf}\rm}
%
%
\def\noblackbox{\overfullrule=0pt}
\hyphenation{anom-aly anom-alies coun-ter-term coun-ter-terms}
\def\inv{^{\raise.15ex\hbox{${\scriptscriptstyle -}$}\kern-.05em 1}}

\def\Dsl{\,\raise.15ex\hbox{/}\mkern-13.5mu D} 
\def\dsl{\raise.15ex\hbox{/}\kern-.57em\partial}

\def\lspace{\ifx\answ\bigans{}\else\qquad\fi}
\def\lbspace{\ifx\answ\bigans{}\else\hskip-.2in\fi} 

\def\boxeqn#1{\vcenter{\vbox{\hrule\hbox{\vrule\kern3pt\vbox{\kern3pt
	\hbox{${\displaystyle #1}$}\kern3pt}\kern3pt\vrule}\hrule}}}
\def\mbox#1#2{\vcenter{\hrule \hbox{\vrule height#2in
		\kern#1in \vrule} \hrule}}  

\def\darr#1{\raise1.5ex\hbox{$\leftrightarrow$}\mkern-16.5mu #1}

\def\roughly#1{\raise.3ex\hbox{$#1$\kern-.75em\lower1ex\hbox{$\sim$}}}

\input amssym

\input epsf

\def\IZ{\relax\ifmmode\mathchoice
{\hbox{\cmss Z\kern-.4em Z}}{\hbox{\cmss Z\kern-.4em Z}} {\lower.9pt\hbox{\cmsss Z\kern-.4em Z}}
{\lower1.2pt\hbox{\cmsss Z\kern-.4em Z}}\else{\cmss Z\kern-.4em Z}\fi}

\newif\ifdraft\draftfalse
\newif\ifinter\interfalse
\ifdraft\draftmode\else\interfalse\fi
\def\journal#1&#2(#3){\unskip, \sl #1\ \bf #2 \rm(19#3) }
\def\andjournal#1&#2(#3){\sl #1~\bf #2 \rm (19#3) }

\def\frac#1#2{{#1\over#2}}

\def\inbar{\,\vrule height1.5ex width.4pt depth0pt}
\def\IC{\relax\hbox{$\inbar\kern-.3em{\rm C}$}}
\def\IR{\relax{\rm I\kern-.18em R}}
\def\IP{\relax{\rm I\kern-.18em P}}

%
%


%
\catcode`\@=11
\def\slash#1{\mathord{\mathpalette\c@ncel{#1}}}
\overfullrule=0pt

\def\CC{{\cal C}}

\def\MM{{\cal M}}

\def\Z{\hbox{$\bb Z$}}

\def\underrel#1\over#2{\mathrel{\mathop{\kern\z@#1}\limits_{#2}}}

\catcode`\@=12


%

\def\mod{{\rm mod}}

\def\exp{{\rm exp}}


\def\[{[}
\def\]{]}

\def\comment#1{ }

%
\def\draftnote#1{\ifdraft{\baselineskip2ex
                 \vbox{\kern1em\hrule\hbox{\vrule\kern1em\vbox{\kern1ex
                 \noindent \underbar{NOTE}: #1
             \vskip1ex}\kern1em\vrule}\hrule}}\fi}
\def\internote#1{\ifinter{\baselineskip2ex
                 \vbox{\kern1em\hrule\hbox{\vrule\kern1em\vbox{\kern1ex
                 \noindent \underbar{Internal Note}: #1
             \vskip1ex}\kern1em\vrule}\hrule}}\fi}

%

%
%

%

\def\inv{^{-1}}


\def\b{\beta}


\def\bb{
\font\tenmsb=msbm10
\font\sevenmsb=msbm7
\font\fivemsb=msbm5
\textfont1=\tenmsb
\scriptfont1=\sevenmsb
\scriptscriptfont1=\fivemsb
}





\def\hat{\widehat}

\def\bar{\overline}
\def\b{\bar}
\def\bsq#1{{{\b{#1}}^{\lower 2.5pt\hbox{$\scriptstyle 2$}}}}
\def\bexp#1#2{{{\b{#1}}^{\lower 2.5pt\hbox{$\scriptstyle #2$}}}}
\def\dotexp#1#2{{{#1}^{\lower 2.5pt\hbox{$\scriptstyle #2$}}}}


\def\Tr{\mathop{\rm Tr}}

\def\rt2{\sqrt{2}}

\def\mod{{\rm mod}}



\def\CC{{\cal C}}

\def\CK{{\cal K}}

\def\CM{{\cal M}}

\def\CO{{\cal O}}

\def\CT{{\cal T}}


\def\1{{\ds 1}}

\def\C{\hbox{$\bb C$}}

\def\Z{\hbox{$\bb Z$}}


\noblackbox

\def\unit{\relax{\rm 1\kern-.26em I}}
\def\nada{\relax{\rm 0\kern-.30em l}}

\def\mod{{\rm \ mod \ }}

\noblackbox
\def\IL{\relax{\rm I\kern-.18em L}}
\def\IH{\relax{\rm I\kern-.18em H}}
\def\IR{\relax{\rm I\kern-.18em R}}
\def\IC{\relax\hbox{$\inbar\kern-.3em{\rm C}$}}
\def\IZ{\relax\ifmmode\mathchoice
{\hbox{\cmss Z\kern-.4em Z}}{\hbox{\cmss Z\kern-.4em Z}} {\lower.9pt\hbox{\cmsss Z\kern-.4em Z}}
{\lower1.2pt\hbox{\cmsss Z\kern-.4em Z}}\else{\cmss Z\kern-.4em Z}\fi}
\def\CM {{\cal M}}

\def\partialslash{\not{\hbox{\kern-2pt $\partial$}}}

\font\manual=manfnt \def\dbend{\lower3.5pt\hbox{\manual\char127}}

\def\IZ{\relax\ifmmode\mathchoice
{\hbox{\cmss Z\kern-.4em Z}}{\hbox{\cmss Z\kern-.4em Z}} {\lower.9pt\hbox{\cmsss Z\kern-.4em Z}}
{\lower1.2pt\hbox{\cmsss Z\kern-.4em Z}}\else{\cmss Z\kern-.4em Z}\fi}
\def\half{{1\over 2}}

\def\bar{\overline}

\def\rt2{\sqrt{2}}
\def\irt2{{1\over\sqrt{2}}}

\def\hat{\widehat}
\def\slashchar#1{\setbox0=\hbox{$#1$}           
   \dimen0=\wd0                                 
   \setbox1=\hbox{/} \dimen1=\wd1               
   \ifdim\dimen0>\dimen1                        
      \rlap{\hbox to \dimen0{\hfil/\hfil}}      
      #1                                        
   \else                                        
      \rlap{\hbox to \dimen1{\hfil$#1$\hfil}}   
      /                                         
   \fi}


\def\figcaption#1#2{\DefWarn#1\xdef#1{Figure~\noexpand\hyperref{}{figure}%
{\the\figno}{\the\figno}}\writedef{#1\leftbracket Figure\noexpand~\xfig#1}%
\medskip\centerline{{\footnotefont\bf Figure~\hyperdef\hypernoname{figure}{\the\figno}{\the\figno}:}  #2 \wrlabeL{#1=#1}}%
\global\advance\figno by1}

%
%

\lref\AffleckAS{
  I.~Affleck, J.~A.~Harvey and E.~Witten,
  ``Instantons and (Super)Symmetry Breaking in (2+1)-Dimensions,''
Nucl.\ Phys.\ B {\bf 206}, 413 (1982).
}

\lref\AharonyBX{
  O.~Aharony, A.~Hanany, K.~A.~Intriligator, N.~Seiberg and M.~J.~Strassler,
  ``Aspects of N=2 supersymmetric gauge theories in three-dimensions,''
Nucl.\ Phys.\ B {\bf 499}, 67 (1997).
[hep-th/9703110].
}

\lref\AharonyGP{
  O.~Aharony,
  ``IR duality in d = 3 N=2 supersymmetric USp(2N(c)) and U(N(c)) gauge theories,''
Phys.\ Lett.\ B {\bf 404}, 71 (1997).
[hep-th/9703215].
}

\lref\AharonyCI{
  O.~Aharony and I.~Shamir,
  ``On $O(N_c)$ d=3 N=2 supersymmetric QCD Theories,''
JHEP {\bf 1112}, 043 (2011).
[arXiv:1109.5081 [hep-th]].
}

\lref\AharonyJZ{
  O.~Aharony, G.~Gur-Ari and R.~Yacoby,
  ``d=3 Bosonic Vector Models Coupled to Chern-Simons Gauge Theories,''
JHEP {\bf 1203}, 037 (2012).
[arXiv:1110.4382 [hep-th]].
}

\lref\AharonyNH{
  O.~Aharony, G.~Gur-Ari and R.~Yacoby,
  ``Correlation Functions of Large N Chern-Simons-Matter Theories and Bosonization in Three Dimensions,''
JHEP {\bf 1212}, 028 (2012).
[arXiv:1207.4593 [hep-th]].
}

\lref\AST{
O.~Aharony, N.~Seiberg and Y.~Tachikawa,
  ``Reading between the lines of four-dimensional gauge theories,''
[arXiv:1305.0318 [hep-th]].
}

\lref\AharonyDHA{
  O.~Aharony, S.~S.~Razamat, N.~Seiberg and B.~Willett,
  ``3d dualities from 4d dualities,''
JHEP {\bf 1307}, 149 (2013).
[arXiv:1305.3924 [hep-th]].
}

\lref\AharonyKMA{
  O.~Aharony, S.~S.~Razamat, N.~Seiberg and B.~Willett,
  ``3$d$ dualities from 4$d$ dualities for orthogonal groups,''
JHEP {\bf 1308}, 099 (2013)
[arXiv:1307.0511 [hep-th]].
}

\lref\AharonyMJS{
  O.~Aharony,
  ``Baryons, monopoles and dualities in Chern-Simons-matter theories,''
JHEP {\bf 1602}, 093 (2016).
[arXiv:1512.00161 [hep-th]].
}

\lref\AlvarezGaumeNF{
  L.~Alvarez-Gaume, S.~Della Pietra and G.~W.~Moore,
  ``Anomalies and Odd Dimensions,''
Annals Phys.\  {\bf 163}, 288 (1985).
}

\lref\AnninosUI{
  D.~Anninos, T.~Hartman and A.~Strominger,
  ``Higher Spin Realization of the dS/CFT Correspondence,''
[arXiv:1108.5735 [hep-th]].
}

\lref\AnninosHIA{
  D.~Anninos, R.~Mahajan, D�.~Radicevic and E.~Shaghoulian,
  ``Chern-Simons-Ghost Theories and de Sitter Space,''
JHEP {\bf 1501}, 074 (2015).
[arXiv:1405.1424 [hep-th]].
}

\lref\AtiyahJF{
  M.~F.~Atiyah, V.~K.~Patodi and I.~M.~Singer,
  ``Spectral Asymmetry in Riemannian Geometry, I,''
  Math.\ Proc.\ Camb.\ Phil.\ Soc.\ {\bf 77} (1975) 43--69.
}

\lref\BanksZN{
  T.~Banks and N.~Seiberg,
  ``Symmetries and Strings in Field Theory and Gravity,''
Phys.\ Rev.\ D {\bf 83}, 084019 (2011).
[arXiv:1011.5120 [hep-th]].
}

\lref\BarkeshliIDA{
  M.~Barkeshli and J.~McGreevy,
  ``Continuous transition between fractional quantum Hall and superfluid states,''
Phys.\ Rev.\ B {\bf 89}, no. 23, 235116 (2014).
}

\lref\BeemMB{
  C.~Beem, T.~Dimofte and S.~Pasquetti,
  ``Holomorphic Blocks in Three Dimensions,''
[arXiv:1211.1986 [hep-th]].
}

\lref\BeniniMF{
  F.~Benini, C.~Closset and S.~Cremonesi,
  ``Comments on 3d Seiberg-like dualities,''
JHEP {\bf 1110}, 075 (2011).
[arXiv:1108.5373 [hep-th]].
}

\lref\BernardXY{
  D.~Bernard,
  ``String Characters From {Kac-Moody} Automorphisms,''
  Nucl.\ Phys.\ B {\bf 288}, 628 (1987).
}

\lref\BhattacharyaZY{
  J.~Bhattacharya, S.~Bhattacharyya, S.~Minwalla and S.~Raju,
  ``Indices for Superconformal Field Theories in 3,5 and 6 Dimensions,''
JHEP {\bf 0802}, 064 (2008).
[arXiv:0801.1435 [hep-th]].
}

\lref\deBoerMP{
  J.~de Boer, K.~Hori, H.~Ooguri and Y.~Oz,
  ``Mirror symmetry in three-dimensional gauge theories, quivers and D-branes,''
Nucl.\ Phys.\ B {\bf 493}, 101 (1997).
[hep-th/9611063].
}

\lref\deBoerKA{
  J.~de Boer, K.~Hori, Y.~Oz and Z.~Yin,
  ``Branes and mirror symmetry in N=2 supersymmetric gauge theories in three-dimensions,''
Nucl.\ Phys.\ B {\bf 502}, 107 (1997).
[hep-th/9702154].
}

\lref\BondersonPLA{
  P.~Bonderson, C.~Nayak and X.~L.~Qi,
  ``A time-reversal invariant topological phase at the surface of a 3D topological insulator,''
J.\ Stat.\ Mech.\  {\bf 2013}, P09016 (2013).
}

\lref\BorokhovIB{
  V.~Borokhov, A.~Kapustin and X.~k.~Wu,
  ``Topological disorder operators in three-dimensional conformal field theory,''
JHEP {\bf 0211}, 049 (2002).
[hep-th/0206054].
}

\lref\BorokhovCG{
  V.~Borokhov, A.~Kapustin and X.~k.~Wu,
  ``Monopole operators and mirror symmetry in three-dimensions,''
JHEP {\bf 0212}, 044 (2002).
[hep-th/0207074].
}

\lref\debult{
  F.~van~de~Bult,
  ``Hyperbolic Hypergeometric Functions,''
University of Amsterdam Ph.D. thesis
}

\lref\Camperi{
  M.~Camperi, F.~Levstein and G.~Zemba,
  ``The Large N Limit Of Chern-simons Gauge Theory,''
  Phys.\ Lett.\ B {\bf 247} (1990) 549.
}

\lref\ChenCD{
  W.~Chen, M.~P.~A.~Fisher and Y.~S.~Wu,
  ``Mott transition in an anyon gas,''
Phys.\ Rev.\ B {\bf 48}, 13749 (1993).
[cond-mat/9301037].
}

\lref\ChenJHA{
  X.~Chen, L.~Fidkowski and A.~Vishwanath,
  ``Symmetry Enforced Non-Abelian Topological Order at the Surface of a Topological Insulator,''
Phys.\ Rev.\ B {\bf 89}, no. 16, 165132 (2014).
[arXiv:1306.3250 [cond-mat.str-el]].
}

\lref\ChengPDN{
  M.~Cheng and C.~Xu,
  ``A series of (2+1)d Stable Self-Dual Interacting Conformal Field Theories,''
[arXiv:1609.02560 [cond-mat.str-el]].
}

\lref\ClossetVG{
  C.~Closset, T.~T.~Dumitrescu, G.~Festuccia, Z.~Komargodski and N.~Seiberg,
  ``Contact Terms, Unitarity, and F-Maximization in Three-Dimensional Superconformal Theories,''
JHEP {\bf 1210}, 053 (2012).
[arXiv:1205.4142 [hep-th]].
}

\lref\ClossetVP{
  C.~Closset, T.~T.~Dumitrescu, G.~Festuccia, Z.~Komargodski and N.~Seiberg,
  ``Comments on Chern-Simons Contact Terms in Three Dimensions,''
JHEP {\bf 1209}, 091 (2012).
[arXiv:1206.5218 [hep-th]].
}

\lref\ClossetRU{
  C.~Closset, T.~T.~Dumitrescu, G.~Festuccia and Z.~Komargodski,
  ``Supersymmetric Field Theories on Three-Manifolds,''
JHEP {\bf 1305}, 017 (2013).
[arXiv:1212.3388 [hep-th]].
}

\lref\CveticXN{
  M.~Cvetic, T.~W.~Grimm and D.~Klevers,
  ``Anomaly Cancellation And Abelian Gauge Symmetries In F-theory,''
JHEP {\bf 1302}, 101 (2013).
[arXiv:1210.6034 [hep-th]].
}

\lref\DaiKQ{
  X.~z.~Dai and D.~S.~Freed,
  ``eta invariants and determinant lines,''
J.\ Math.\ Phys.\  {\bf 35}, 5155 (1994), Erratum: [J.\ Math.\ Phys.\  {\bf 42}, 2343 (2001)].
[hep-th/9405012].
}

\lref\DasguptaZZ{
  C.~Dasgupta and B.~I.~Halperin,
  ``Phase Transition in a Lattice Model of Superconductivity,''
Phys.\ Rev.\ Lett.\  {\bf 47}, 1556 (1981).
}

\lref\DaviesUW{
  N.~M.~Davies, T.~J.~Hollowood, V.~V.~Khoze and M.~P.~Mattis,
  ``Gluino condensate and magnetic monopoles in supersymmetric gluodynamics,''
Nucl.\ Phys.\ B {\bf 559}, 123 (1999).
[hep-th/9905015].
}

\lref\DaviesNW{
  N.~M.~Davies, T.~J.~Hollowood and V.~V.~Khoze,
  ``Monopoles, affine algebras and the gluino condensate,''
J.\ Math.\ Phys.\  {\bf 44}, 3640 (2003).
[hep-th/0006011].
}

\lref\DimoftePY{
  T.~Dimofte, D.~Gaiotto and S.~Gukov,
  ``3-Manifolds and 3d Indices,''
[arXiv:1112.5179 [hep-th]].
}

\lref\DolanQI{
  F.~A.~Dolan and H.~Osborn,
  ``Applications of the Superconformal Index for Protected Operators and q-Hypergeometric Identities to N=1 Dual Theories,''
Nucl.\ Phys.\ B {\bf 818}, 137 (2009).
[arXiv:0801.4947 [hep-th]].
}

\lref\DolanRP{
  F.~A.~H.~Dolan, V.~P.~Spiridonov and G.~S.~Vartanov,
  ``From 4d superconformal indices to 3d partition functions,''
Phys.\ Lett.\ B {\bf 704}, 234 (2011).
[arXiv:1104.1787 [hep-th]].
}

\lref\DouglasEX{
  M.~R.~Douglas,
  ``Chern-Simons-Witten theory as a topological Fermi liquid,''
[hep-th/9403119].
}

\lref\EagerHX{
  R.~Eager, J.~Schmude and Y.~Tachikawa,
  ``Superconformal Indices, Sasaki-Einstein Manifolds, and Cyclic Homologies,''
[arXiv:1207.0573 [hep-th]].
}

\lref\ElitzurFH{
  S.~Elitzur, A.~Giveon and D.~Kutasov,
  ``Branes and N=1 duality in string theory,''
Phys.\ Lett.\ B {\bf 400}, 269 (1997).
[hep-th/9702014].
}

\lref\ElitzurHC{
  S.~Elitzur, A.~Giveon, D.~Kutasov, E.~Rabinovici and A.~Schwimmer,
  ``Brane dynamics and N=1 supersymmetric gauge theory,''
Nucl.\ Phys.\ B {\bf 505}, 202 (1997).
[hep-th/9704104].
}

\lref\slthreeZ{
  J.~Felder, A.~Varchenko,
  ``The elliptic gamma function and $SL(3,Z) \times Z^3$,'' $\;\;$
[arXiv:math/0001184].
}

\lref\FestucciaWS{
  G.~Festuccia and N.~Seiberg,
  ``Rigid Supersymmetric Theories in Curved Superspace,''
JHEP {\bf 1106}, 114 (2011).
[arXiv:1105.0689 [hep-th]].
}

\lref\FidkowskiJUA{
  L.~Fidkowski, X.~Chen and A.~Vishwanath,
  ``Non-Abelian Topological Order on the Surface of a 3D Topological Superconductor from an Exactly Solved Model,''
Phys.\ Rev.\ X {\bf 3}, no. 4, 041016 (2013).
[arXiv:1305.5851 [cond-mat.str-el]].
}

\lref\FradkinTT{
  E.~H.~Fradkin and F.~A.~Schaposnik,
  ``The Fermion - boson mapping in three-dimensional quantum field theory,''
Phys.\ Lett.\ B {\bf 338}, 253 (1994).
[hep-th/9407182].
}

\lref\GaddeEN{
  A.~Gadde, L.~Rastelli, S.~S.~Razamat and W.~Yan,
  ``On the Superconformal Index of N=1 IR Fixed Points: A Holographic Check,''
JHEP {\bf 1103}, 041 (2011).
[arXiv:1011.5278 [hep-th]].
}

\lref\GaddeIA{
  A.~Gadde and W.~Yan,
  ``Reducing the 4d Index to the $S^3$ Partition Function,''
JHEP {\bf 1212}, 003 (2012).
[arXiv:1104.2592 [hep-th]].
}

\lref\GaddeDDA{
  A.~Gadde and S.~Gukov,
  ``2d Index and Surface operators,''
[arXiv:1305.0266 [hep-th]].
}

\lref\GaiottoAK{
  D.~Gaiotto and E.~Witten,
  ``S-Duality of Boundary Conditions In N=4 Super Yang-Mills Theory,''
Adv.\ Theor.\ Math.\ Phys.\  {\bf 13}, no. 3, 721 (2009).
[arXiv:0807.3720 [hep-th]].
}

\lref\GaiottoBE{
  D.~Gaiotto, G.~W.~Moore and A.~Neitzke,
  ``Framed BPS States,''
[arXiv:1006.0146 [hep-th]].
}

\lref\GaiottoKFA{
  D.~Gaiotto, A.~Kapustin, N.~Seiberg and B.~Willett,
  ``Generalized Global Symmetries,''
JHEP {\bf 1502}, 172 (2015).
[arXiv:1412.5148 [hep-th]].
}

\lref\GeraedtsPVA{
  S.~D.~Geraedts, M.~P.~Zaletel, R.~S.~K.~Mong, M.~A.~Metlitski, A.~Vishwanath and O.~I.~Motrunich,
  ``The half-filled Landau level: the case for Dirac composite fermions,''
Science {\bf 352}, 197 (2016).
[arXiv:1508.04140 [cond-mat.str-el]].
}

\lref\GiombiKC{
  S.~Giombi, S.~Minwalla, S.~Prakash, S.~P.~Trivedi, S.~R.~Wadia and X.~Yin,
  ``Chern-Simons Theory with Vector Fermion Matter,''
Eur.\ Phys.\ J.\ C {\bf 72}, 2112 (2012).
[arXiv:1110.4386 [hep-th]].
}

\lref\GiveonSR{
  A.~Giveon and D.~Kutasov,
  ``Brane dynamics and gauge theory,''
Rev.\ Mod.\ Phys.\  {\bf 71}, 983 (1999).
[hep-th/9802067].
}

\lref\GiveonZN{
  A.~Giveon and D.~Kutasov,
  ``Seiberg Duality in Chern-Simons Theory,''
Nucl.\ Phys.\ B {\bf 812}, 1 (2009).
[arXiv:0808.0360 [hep-th]].
}

\lref\GoddardVK{
  P.~Goddard, A.~Kent and D.~I.~Olive,
  ``Virasoro Algebras and Coset Space Models,''
Phys.\ Lett.\ B {\bf 152}, 88 (1985).
}

\lref\GreenDA{
  D.~Green, Z.~Komargodski, N.~Seiberg, Y.~Tachikawa and B.~Wecht,
  ``Exactly Marginal Deformations and Global Symmetries,''
JHEP {\bf 1006}, 106 (2010).
[arXiv:1005.3546 [hep-th]].
}

\lref\GurPCA{
  G.~Gur-Ari and R.~Yacoby,
  ``Three Dimensional Bosonization From Supersymmetry,''
JHEP {\bf 1511}, 013 (2015).
[arXiv:1507.04378 [hep-th]].
}

\lref\HalperinMH{
  B.~I.~Halperin, P.~A.~Lee and N.~Read,
  ``Theory of the half filled Landau level,''
Phys.\ Rev.\ B {\bf 47}, 7312 (1993).
}

\lref\HamaEA{
  N.~Hama, K.~Hosomichi and S.~Lee,
  ``SUSY Gauge Theories on Squashed Three-Spheres,''
JHEP {\bf 1105}, 014 (2011).
[arXiv:1102.4716 [hep-th]].
}

\lref\Hasegawa{
K.~Hasegawa,
  ``Spin Module Versions of Weyl's Reciprocity Theorem for Classical Kac-Moody Lie Algebras - An Application to Branching Rule Duality,''
Publ.\ Res.\ Inst.\ Math.\ Sci.\ {\bf 25}, 741-828 (1989).
}

\lref\HoriDK{
  K.~Hori and D.~Tong,
  ``Aspects of Non-Abelian Gauge Dynamics in Two-Dimensional N=(2,2) Theories,''
JHEP {\bf 0705}, 079 (2007).
[hep-th/0609032].
}

\lref\HoriPD{
  K.~Hori,
  ``Duality In Two-Dimensional (2,2) Supersymmetric Non-Abelian Gauge Theories,''
[arXiv:1104.2853 [hep-th]].
}

\lref\HsinBLU{
  P.~S.~Hsin and N.~Seiberg,
  ``Level/rank Duality and Chern-Simons-Matter Theories,''
JHEP {\bf 1609}, 095 (2016).
[arXiv:1607.07457 [hep-th]].
}

\lref\HwangQT{
  C.~Hwang, H.~Kim, K.~-J.~Park and J.~Park,
  ``Index computation for 3d Chern-Simons matter theory: test of Seiberg-like duality,''
JHEP {\bf 1109}, 037 (2011).
[arXiv:1107.4942 [hep-th]].
}

\lref\HwangHT{
  C.~Hwang, K.~-J.~Park and J.~Park,
  ``Evidence for Aharony duality for orthogonal gauge groups,''
JHEP {\bf 1111}, 011 (2011).
[arXiv:1109.2828 [hep-th]].
}

\lref\HwangJH{
  C.~Hwang, H.~-C.~Kim and J.~Park,
  ``Factorization of the 3d superconformal index,''
[arXiv:1211.6023 [hep-th]].
}

\lref\ImamuraSU{
  Y.~Imamura and S.~Yokoyama,
  ``Index for three dimensional superconformal field theories with general R-charge assignments,''
JHEP {\bf 1104}, 007 (2011).
[arXiv:1101.0557 [hep-th]].
}

\lref\ImamuraUW{
  Y.~Imamura,
 ``Relation between the 4d superconformal index and the $S^3$ partition function,''
JHEP {\bf 1109}, 133 (2011).
[arXiv:1104.4482 [hep-th]].
}

\lref\ImamuraWG{
  Y.~Imamura and D.~Yokoyama,
 ``N=2 supersymmetric theories on squashed three-sphere,''
Phys.\ Rev.\ D {\bf 85}, 025015 (2012).
[arXiv:1109.4734 [hep-th]].
}

\lref\ImamuraRQ{
  Y.~Imamura and D.~Yokoyama,
 ``$S^3/Z_n$ partition function and dualities,''
JHEP {\bf 1211}, 122 (2012).
[arXiv:1208.1404 [hep-th]].
}

\lref\IntriligatorID{
  K.~A.~Intriligator and N.~Seiberg,
  ``Duality, monopoles, dyons, confinement and oblique confinement in supersymmetric SO(N(c)) gauge theories,''
Nucl.\ Phys.\ B {\bf 444}, 125 (1995).
[hep-th/9503179].
}

\lref\IntriligatorNE{
  K.~A.~Intriligator and P.~Pouliot,
  ``Exact superpotentials, quantum vacua and duality in supersymmetric SP(N(c)) gauge theories,''
Phys.\ Lett.\ B {\bf 353}, 471 (1995).
[hep-th/9505006].
}

\lref\IntriligatorER{
  K.~A.~Intriligator and N.~Seiberg,
  ``Phases of N=1 supersymmetric gauge theories and electric - magnetic triality,''
In *Los Angeles 1995, Future perspectives in string theory* 270-282.
[hep-th/9506084].
}

\lref\IntriligatorAU{
  K.~A.~Intriligator and N.~Seiberg,
  ``Lectures on supersymmetric gauge theories and electric - magnetic duality,''
Nucl.\ Phys.\ Proc.\ Suppl.\  {\bf 45BC}, 1 (1996).
[hep-th/9509066].
}

\lref\IntriligatorEX{
  K.~A.~Intriligator and N.~Seiberg,
  ``Mirror symmetry in three-dimensional gauge theories,''
Phys.\ Lett.\ B {\bf 387}, 513 (1996).
[hep-th/9607207].
}

\lref\newIS{
K.~Intriligator and N.~Seiberg,
  ``Aspects of 3d N=2 Chern-Simons-Matter Theories,''
[arXiv:1305.1633 [hep-th]].
}

\lref\IvanovFN{
   E.~A.~Ivanov,
   ``Chern-Simons matter systems with manifest N=2 supersymmetry,''
Phys.\ Lett.\ B {\bf 268}, 203 (1991).
}

\lref\JafferisUN{
  D.~L.~Jafferis,
  ``The Exact Superconformal R-Symmetry Extremizes Z,''
JHEP {\bf 1205}, 159 (2012).
[arXiv:1012.3210 [hep-th]].
}

\lref\JafferisZI{
  D.~L.~Jafferis, I.~R.~Klebanov, S.~S.~Pufu and B.~R.~Safdi,
  ``Towards the F-Theorem: N=2 Field Theories on the Three-Sphere,''
JHEP {\bf 1106}, 102 (2011).
[arXiv:1103.1181 [hep-th]].
}

\lref\JainTX{
  J.~K.~Jain,
  ``Composite fermion approach for the fractional quantum Hall effect,''
Phys.\ Rev.\ Lett.\  {\bf 63}, 199 (1989).
}

\lref\JainGZA{
  S.~Jain, S.~Minwalla and S.~Yokoyama,
  ``Chern Simons duality with a fundamental boson and fermion,''
JHEP {\bf 1311}, 037 (2013).
[arXiv:1305.7235 [hep-th]].
}

\lref\KachruRMA{
  S.~Kachru, M.~Mulligan, G.~Torroba and H.~Wang,
  ``Mirror symmetry and the half-filled Landau level,''
Phys.\ Rev.\ B {\bf 92}, 235105 (2015).
[arXiv:1506.01376 [cond-mat.str-el]].
}

\lref\KachruRUI{
  S.~Kachru, M.~Mulligan, G.~Torroba and H.~Wang,
  ``Bosonization and Mirror Symmetry,''
Phys.\ Rev.\ D {\bf 94}, no. 8, 085009 (2016).
[arXiv:1608.05077 [hep-th]].
}

\lref\KachruAON{
  S.~Kachru, M.~Mulligan, G.~Torroba and H.~Wang,
  ``The many faces of mirror symmetry,''
[arXiv:1609.02149 [hep-th]].
}

\lref\KajantieVY{
  K.~Kajantie, M.~Laine, T.~Neuhaus, A.~Rajantie and K.~Rummukainen,
  ``Duality and scaling in three-dimensional scalar electrodynamics,''
Nucl.\ Phys.\ B {\bf 699}, 632 (2004).
[hep-lat/0402021].
}

\lref\KapustinHA{
  A.~Kapustin and M.~J.~Strassler,
  ``On mirror symmetry in three-dimensional Abelian gauge theories,''
JHEP {\bf 9904}, 021 (1999).
[hep-th/9902033].
}

\lref\KapustinPY{
  A.~Kapustin,
  ``Wilson-'t Hooft operators in four-dimensional gauge theories and S-duality,''
Phys.\ Rev.\ D {\bf 74}, 025005 (2006).
[hep-th/0501015].
}

\lref\KapustinKZ{
  A.~Kapustin, B.~Willett and I.~Yaakov,
  ``Exact Results for Wilson Loops in Superconformal Chern-Simons Theories with Matter,''
JHEP {\bf 1003}, 089 (2010).
[arXiv:0909.4559 [hep-th]].
}

\lref\KapustinSim{
A.~Kapustin,  2010 Simons Workshop talk, a video of this talk can be found at
{\tt
http://media.scgp.stonybrook.edu/video/video.php?f=20110810\_1\_qtp.mp4}
}

\lref\KapustinXQ{
  A.~Kapustin, B.~Willett and I.~Yaakov,
  ``Nonperturbative Tests of Three-Dimensional Dualities,''
JHEP {\bf 1010}, 013 (2010).
[arXiv:1003.5694 [hep-th]].
}

\lref\KapustinGH{
  A.~Kapustin,
  ``Seiberg-like duality in three dimensions for orthogonal gauge groups,''
arXiv:1104.0466 [hep-th].
}

\lref\KapustinJM{
  A.~Kapustin and B.~Willett,
  ``Generalized Superconformal Index for Three Dimensional Field Theories,''
[arXiv:1106.2484 [hep-th]].
}

\lref\KapustinVZ{
  A.~Kapustin, H.~Kim and J.~Park,
  ``Dualities for 3d Theories with Tensor Matter,''
JHEP {\bf 1112}, 087 (2011).
[arXiv:1110.2547 [hep-th]].
}

\lref\KapustinGUA{
  A.~Kapustin and N.~Seiberg,
  ``Coupling a QFT to a TQFT and Duality,''
JHEP {\bf 1404}, 001 (2014).
[arXiv:1401.0740 [hep-th]].
}

\lref\KarchUX{
  A.~Karch,
  ``Seiberg duality in three-dimensions,''
Phys.\ Lett.\ B {\bf 405}, 79 (1997).
[hep-th/9703172].
}

\lref\KarchSXI{
  A.~Karch and D.~Tong,
  ``Particle-Vortex Duality from 3d Bosonization,''
[arXiv: 1606.01893 [hep-th]].
}

\lref\KimWB{
  S.~Kim,
  ``The Complete superconformal index for N=6 Chern-Simons theory,''
Nucl.\ Phys.\ B {\bf 821}, 241 (2009), [Erratum-ibid.\ B {\bf 864}, 884 (2012)].
[arXiv:0903.4172 [hep-th]].
}

\lref\KimCMA{
  H.~Kim and J.~Park,
  ``Aharony Dualities for 3d Theories with Adjoint Matter,''
[arXiv:1302.3645 [hep-th]].
}

\lref\KinneyEJ{
  J.~Kinney, J.~M.~Maldacena, S.~Minwalla and S.~Raju,
  ``An Index for 4 dimensional super conformal theories,''
Commun.\ Math.\ Phys.\  {\bf 275}, 209 (2007).
[hep-th/0510251].
}

\lref\KlebanovJA{
  I.~R.~Klebanov and A.~M.~Polyakov,
  ``AdS dual of the critical O(N) vector model,''
Phys.\ Lett.\ B {\bf 550}, 213 (2002).
[hep-th/0210114].
}

\lref\KrattenthalerDA{
  C.~Krattenthaler, V.~P.~Spiridonov, G.~S.~Vartanov,
  ``Superconformal indices of three-dimensional theories related by mirror symmetry,''
JHEP {\bf 1106}, 008 (2011).
[arXiv:1103.4075 [hep-th]].
}

\lref\McG{
  S. M. Kravec, J. McGreevy, and B. Swingle,
  ``All-Fermion Electrodynamics And Fermion Number Anomaly Inflow,''
arXiv:1409.8339.
}

\lref\KutasovVE{
  D.~Kutasov,
  ``A Comment on duality in N=1 supersymmetric nonAbelian gauge theories,''
Phys.\ Lett.\ B {\bf 351}, 230 (1995).
[hep-th/9503086].
}

\lref\KutasovNP{
  D.~Kutasov and A.~Schwimmer,
  ``On duality in supersymmetric Yang-Mills theory,''
Phys.\ Lett.\ B {\bf 354}, 315 (1995).
[hep-th/9505004].
}

\lref\KutasovSS{
  D.~Kutasov, A.~Schwimmer and N.~Seiberg,
  ``Chiral rings, singularity theory and electric - magnetic duality,''
Nucl.\ Phys.\ B {\bf 459}, 455 (1996).
[hep-th/9510222].
}

\lref\LeeVP{
  K.~-M.~Lee and P.~Yi,
  ``Monopoles and instantons on partially compactified D-branes,''
Phys.\ Rev.\ D {\bf 56}, 3711 (1997).
[hep-th/9702107].
}

\lref\LeeVU{
  K.~-M.~Lee,
  ``Instantons and magnetic monopoles on R**3 x S**1 with arbitrary simple gauge groups,''
Phys.\ Lett.\ B {\bf 426}, 323 (1998).
[hep-th/9802012].
}

\lref\MaldacenaSS{
  J.~M.~Maldacena, G.~W.~Moore and N.~Seiberg,
  ``D-brane charges in five-brane backgrounds,''
JHEP {\bf 0110}, 005 (2001).
[hep-th/0108152].
}

\lref\MetlitskiEKA{
  M.~A.~Metlitski and A.~Vishwanath,
  ``Particle-vortex duality of 2d Dirac fermion from electric-magnetic duality of 3d topological insulators,''
[arXiv:1505.05142 [cond-mat.str-el]].
}

\lref\MetlitskiBPA{
  M.~A.~Metlitski, C.~L.~Kane and M.~P.~A.~Fisher,
  ``Symmetry-respecting topologically ordered surface phase of three-dimensional electron topological insulators,''
Phys.\ Rev.\ B {\bf 92}, no. 12, 125111 (2015).
}

\lref\MetlitskiYQA{
  M.~A.~Metlitski,
  ``$S$-duality of $u(1)$ gauge theory with $\theta =\pi$ on non-orientable manifolds: Applications to topological insulators and superconductors,''
[arXiv:1510.05663 [hep-th]].
}

\lref\MetlitskiDHT{
  M.~A.~Metlitski, A.~Vishwanath and C.~Xu,
  ``Duality and bosonization of (2+1)d Majorana fermions,''
  arXiv:1611.05049 [cond-mat.str-el].
}

\lref\MlawerUV{
  E.~J.~Mlawer, S.~G.~Naculich, H.~A.~Riggs and H.~J.~Schnitzer,
  ``Group level duality of WZW fusion coefficients and Chern-Simons link observables,''
Nucl.\ Phys.\ B {\bf 352}, 863 (1991).
}

\lref\MSN{
G.~W.~Moore and N.~Seiberg,
  ``Naturality in Conformal Field Theory,''
  Nucl.\ Phys.\ B {\bf 313}, 16 (1989).}

\lref\MooreYH{
  G.~W.~Moore and N.~Seiberg,
  ``Taming the Conformal Zoo,''
Phys.\ Lett.\ B {\bf 220}, 422 (1989).
}

\lref\MoritaCS{
  T.~Morita and V.~Niarchos,
  ``F-theorem, duality and SUSY breaking in one-adjoint Chern-Simons-Matter theories,''
Nucl.\ Phys.\ B {\bf 858}, 84 (2012).
[arXiv:1108.4963 [hep-th]].
}

\lref\MrossIDY{
  D.~F.~Mross, J.~Alicea and O.~I.~Motrunich,
  ``Explicit derivation of duality between a free Dirac cone and quantum electrodynamics in (2+1) dimensions,''
[arXiv:1510.08455 [cond-mat.str-el]].
}

\lref\MulliganGLM{
  M.~Mulligan, S.~Raghu and M.~P.~A.~Fisher,
  ``Emergent particle-hole symmetry in the half-filled Landau level,''
[arXiv:1603.05656 [cond-mat.str-el]].
}

\lref\MuruganZAL{
  J.~Murugan and H.~Nastase,
  ``Particle-vortex duality in topological insulators and superconductors,''
[arXiv:1606.01912 [hep-th]].
}

\lref\NaculichPA{
  S.~G.~Naculich, H.~A.~Riggs and H.~J.~Schnitzer,
  ``Group Level Duality in {WZW} Models and {Chern-Simons} Theory,''
Phys.\ Lett.\ B {\bf 246}, 417 (1990).
}

\lref\NaculichNC{
  S.~G.~Naculich and H.~J.~Schnitzer,
  ``Level-rank duality of the U(N) WZW model, Chern-Simons theory, and 2-D qYM theory,''
JHEP {\bf 0706}, 023 (2007).
[hep-th/0703089 [HEP-TH]].
}

\lref\NakaharaNW{
  M.~Nakahara,
  ``Geometry, topology and physics,''
Boca Raton, USA: Taylor and Francis (2003) 573 p.
}

\lref\NakanishiHJ{
  T.~Nakanishi and A.~Tsuchiya,
  ``Level rank duality of WZW models in conformal field theory,''
Commun.\ Math.\ Phys.\  {\bf 144}, 351 (1992).
}

\lref\NguyenZN{
  A.~K.~Nguyen and A.~Sudbo,
  ``Topological phase fluctuations, amplitude fluctuations, and criticality in extreme type II superconductors,''
Phys.\ Rev.\ B {\bf 60}, 15307 (1999).
[cond-mat/9907385].
}

\lref\NiarchosJB{
  V.~Niarchos,
  ``Seiberg Duality in Chern-Simons Theories with Fundamental and Adjoint Matter,''
JHEP {\bf 0811}, 001 (2008).
[arXiv:0808.2771 [hep-th]].
}

\lref\NiarchosAA{
  V.~Niarchos,
  ``R-charges, Chiral Rings and RG Flows in Supersymmetric Chern-Simons-Matter Theories,''
JHEP {\bf 0905}, 054 (2009).
[arXiv:0903.0435 [hep-th]].
}

\lref\NiarchosAH{
  V.~Niarchos,
  ``Seiberg dualities and the 3d/4d connection,''
JHEP {\bf 1207}, 075 (2012).
[arXiv:1205.2086 [hep-th]].
}

\lref\NiemiRQ{
  A.~J.~Niemi and G.~W.~Semenoff,
  ``Axial Anomaly Induced Fermion Fractionization and Effective Gauge Theory Actions in Odd Dimensional Space-Times,''
Phys.\ Rev.\ Lett.\  {\bf 51}, 2077 (1983).
}

\lref\VOstrik{
  V.~Ostrik and M.~Sun,
  ``Level-Rank Duality Via Tensor Categories,''
Comm. Math. Phys. 326 (2014) 49-61.
[arXiv:1208.5131 [math-ph]].
}

\lref\ParkWTA{
  J.~Park and K.~J.~Park,
  ``Seiberg-like Dualities for 3d N=2 Theories with SU(N) gauge group,''
JHEP {\bf 1310}, 198 (2013).
[arXiv:1305.6280 [hep-th]].
}

\lref\PaulyAMA{
  C.~Pauly,
  ``Strange duality revisited,''
Math.\ Res.\ Lett.\  {\bf 21}, 1353 (2014).
}

\lref\PeskinKP{
  M.~E.~Peskin,
  ``Mandelstam 't Hooft Duality in Abelian Lattice Models,''
Annals Phys.\  {\bf 113}, 122 (1978).
}

\lref\PolyakovFU{
  A.~M.~Polyakov,
  ``Quark Confinement and Topology of Gauge Groups,''
Nucl.\ Phys.\ B {\bf 120}, 429 (1977).
}

\lref\PolyakovMD{
  A.~M.~Polyakov,
  ``Fermi-Bose Transmutations Induced by Gauge Fields,''
Mod.\ Phys.\ Lett.\ A {\bf 3}, 325 (1988).
}

\lref\PotterCDN{
  A.~C.~Potter, M.~Serbyn and A.~Vishwanath,
  ``Thermoelectric transport signatures of Dirac composite fermions in the half-filled Landau level,''
[arXiv:1512.06852 [cond-mat.str-el]].
}

\lref\RadicevicYLA{
  D.~Radicevic,
  ``Disorder Operators in Chern-Simons-Fermion Theories,''
JHEP {\bf 1603}, 131 (2016).
[arXiv:1511.01902 [hep-th]].
}

\lref\RazamatUV{
  S.~S.~Razamat,
  ``On a modular property of N=2 superconformal theories in four dimensions,''
JHEP {\bf 1210}, 191 (2012).
[arXiv:1208.5056 [hep-th]].
}

\lref\RedlichDV{
  A.~N.~Redlich,
  ``Parity Violation and Gauge Noninvariance of the Effective Gauge Field Action in Three-Dimensions,''
Phys.\ Rev.\ D {\bf 29}, 2366 (1984).
}

\lref\Rehren{
  K.-H.~Rehren,
  ``Algebraic Conformal QFT'',
3rd Meeting of the French-Italian Research Team on Noncommutative Geometry and Quantum Physics Vietri sul Mare, 2009.
}

\lref\RomelsbergerEG{
  C.~Romelsberger,
  ``Counting chiral primaries in N = 1, d=4 superconformal field theories,''
Nucl.\ Phys.\ B {\bf 747}, 329 (2006).
[hep-th/0510060].
}

\lref\RoscherWOX{
  D.~Roscher, E.~Torres and P.~Strack,
  ``Dual QED$_3$ at "$N_F = 1/2$" is an interacting CFT in the infrared,''
[arXiv:1605.05347 [cond-mat.str-el]].
}

\lref\SafdiRE{
  B.~R.~Safdi, I.~R.~Klebanov and J.~Lee,
  ``A Crack in the Conformal Window,''
[arXiv:1212.4502 [hep-th]].
}

\lref\SeibergBZ{
  N.~Seiberg,
  ``Exact results on the space of vacua of four-dimensional SUSY gauge theories,''
Phys.\ Rev.\ D {\bf 49}, 6857 (1994).
[hep-th/9402044].
}

\lref\SeibergPQ{
  N.~Seiberg,
  ``Electric - magnetic duality in supersymmetric nonAbelian gauge theories,''
Nucl.\ Phys.\ B {\bf 435}, 129 (1995).
[hep-th/9411149].
}

\lref\SeibergNZ{
  N.~Seiberg and E.~Witten,
  ``Gauge dynamics and compactification to three-dimensions,''
In *Saclay 1996, The mathematical beauty of physics* 333-366.
[hep-th/9607163].
}

\lref\SeibergQD{
  N.~Seiberg,
  ``Modifying the Sum Over Topological Sectors and Constraints on Supergravity,''
JHEP {\bf 1007}, 070 (2010).
[arXiv:1005.0002 [hep-th]].
}

\lref\SeibergRSG{
  N.~Seiberg and E.~Witten,
  ``Gapped Boundary Phases of Topological Insulators via Weak Coupling,''
[arXiv:1602.04251 [cond-mat.str-el]].
}

\lref\SeibergGMD{
  N.~Seiberg, T.~Senthil, C.~Wang and E.~Witten,
  ``A Duality Web in 2+1 Dimensions and Condensed Matter Physics,''
Annals Phys.\  {\bf 374}, 395 (2016).
[arXiv:1606.01989 [hep-th]].
}

\lref\SenthilJK{
  T.~Senthil and M.~P.~A.~Fisher,
  ``Competing orders, non-linear sigma models, and topological terms in quantum magnets,''
Phys.\ Rev.\ B {\bf 74}, 064405 (2006).
[cond-mat/0510459].
}

\lref\SezginPT{
  E.~Sezgin and P.~Sundell,
  ``Holography in 4D (super) higher spin theories and a test via cubic scalar couplings,''
JHEP {\bf 0507}, 044 (2005).
[hep-th/0305040].
}

\lref\ShajiIS{
  N.~Shaji, R.~Shankar and M.~Sivakumar,
  ``On Bose-fermi Equivalence in a U(1) Gauge Theory With {Chern-Simons} Action,''
Mod.\ Phys.\ Lett.\ A {\bf 5}, 593 (1990).
}

\lref\Shamirthesis{
  I.~Shamir,
  ``Aspects of three dimensional Seiberg duality,''
  M.~Sc. thesis submitted to the Weizmann Institute of Science, April 2010.
}

\lref\ShenkerZF{
  S.~H.~Shenker and X.~Yin,
  ``Vector Models in the Singlet Sector at Finite Temperature,''
[arXiv:1109.3519 [hep-th]].
}

\lref\SonXQA{
  D.~T.~Son,
  ``Is the Composite Fermion a Dirac Particle?,''
Phys.\ Rev.\ X {\bf 5}, no. 3, 031027 (2015).
[arXiv:1502.03446 [cond-mat.mes-hall]].
}

\lref\SpiridonovZR{
  V.~P.~Spiridonov and G.~S.~Vartanov,
  ``Superconformal indices for N = 1 theories with multiple duals,''
Nucl.\ Phys.\ B {\bf 824}, 192 (2010).
[arXiv:0811.1909 [hep-th]].
}

\lref\SpiridonovZA{
  V.~P.~Spiridonov and G.~S.~Vartanov,
  ``Elliptic Hypergeometry of Supersymmetric Dualities,''
Commun.\ Math.\ Phys.\  {\bf 304}, 797 (2011).
[arXiv:0910.5944 [hep-th]].
}

\lref\SpiridonovHF{
  V.~P.~Spiridonov and G.~S.~Vartanov,
  ``Elliptic hypergeometry of supersymmetric dualities II. Orthogonal groups, knots, and vortices,''
[arXiv:1107.5788 [hep-th]].
}

\lref\SpiridonovWW{
  V.~P.~Spiridonov and G.~S.~Vartanov,
  ``Elliptic hypergeometric integrals and 't Hooft anomaly matching conditions,''
JHEP {\bf 1206}, 016 (2012).
[arXiv:1203.5677 [hep-th]].
}

\lref\StrasslerFE{
  M.~J.~Strassler,
  ``Duality, phases, spinors and monopoles in $SO(N)$ and $spin(N)$ gauge theories,''
JHEP {\bf 9809}, 017 (1998).
[hep-th/9709081].
}

\lref\VasilievVF{
  M.~A.~Vasiliev,
  ``Holography, Unfolding and Higher-Spin Theory,''
J.\ Phys.\ A {\bf 46}, 214013 (2013).
[arXiv:1203.5554 [hep-th]].
}

\lref\VerstegenAT{
  D.~Verstegen,
  ``Conformal embeddings, rank-level duality and exceptional modular invariants,''
Commun.\ Math.\ Phys.\  {\bf 137}, 567 (1991).
}

\lref\WangUKY{
  C.~Wang, A.~C.~Potter and T.~Senthil,
  ``Gapped symmetry preserving surface state for the electron topological insulator,''
Phys.\ Rev.\ B {\bf 88}, no. 11, 115137 (2013).
[arXiv:1306.3223 [cond-mat.str-el]].
}

\lref\MPS{
  C. Wang, A. C. Potter, and T. Senthil,
  ``Classification Of Interacting Electronic Topological Insulators In Three Dimensions,''
Science {\bf 343} (2014) 629,
[arXiv:1306.3238].
}

\lref\WangLCA{
  C.~Wang and T.~Senthil,
  ``Interacting fermionic topological insulators/superconductors in three dimensions,''
Phys.\ Rev.\ B {\bf 89}, no. 19, 195124 (2014), Erratum: [Phys.\ Rev.\ B {\bf 91}, no. 23, 239902 (2015)].
[arXiv:1401.1142 [cond-mat.str-el]].
}

\lref\WangQMT{
  C.~Wang and T.~Senthil,
  ``Dual Dirac Liquid on the Surface of the Electron Topological Insulator,''
Phys.\ Rev.\ X {\bf 5}, no. 4, 041031 (2015). [arXiv:1505.05141 [cond-mat.str-el]].
}

\lref\WangFQL{
  C.~Wang and T.~Senthil,
  ``Half-filled Landau level, topological insulator surfaces, and three-dimensional quantum spin liquids,''
Phys.\ Rev.\ B {\bf 93}, no. 8, 085110 (2016). [arXiv:1507.08290 [cond-mat.st-el]].
}

\lref\WangGQJ{
  C.~Wang and T.~Senthil,
  ``Composite fermi liquids in the lowest Landau level,''
[arXiv:1604.06807 [cond-mat.str-el]].
}

\lref\WangCTO{
  C.~Wang and T.~Senthil,
  ``Time-Reversal Symmetric $U(1)$ Quantum Spin Liquids,''
Phys.\ Rev.\ X {\bf 6}, no. 1, 011034 (2016).
}

\lref\WilczekDU{
  F.~Wilczek,
  ``Magnetic Flux, Angular Momentum, and Statistics,''
Phys.\ Rev.\ Lett.\  {\bf 48}, 1144 (1982).
}

\lref\WillettGP{
  B.~Willett and I.~Yaakov,
  ``N=2 Dualities and Z Extremization in Three Dimensions,''
arXiv:1104.0487 [hep-th].
}

\lref\WittenHF{
  E.~Witten,
  ``Quantum Field Theory and the Jones Polynomial,''
Commun.\ Math.\ Phys.\  {\bf 121}, 351 (1989).
}

\lref\WittenXI{
  E.~Witten,
  ``The Verlinde algebra and the cohomology of the Grassmannian,''
In *Cambridge 1993, Geometry, topology, and physics* 357-422.
[hep-th/9312104].
}

\lref\WittenGF{
  E.~Witten,
  ``On S duality in Abelian gauge theory,''
Selecta Math.\  {\bf 1}, 383 (1995).
[hep-th/9505186].
}

\lref\WittenDS{
  E.~Witten,
  ``Supersymmetric index of three-dimensional gauge theory,''
In *Shifman, M.A. (ed.): The many faces of the superworld* 156-184.
[hep-th/9903005].
}

\lref\WittenYA{
  E.~Witten,
  ``SL(2,Z) action on three-dimensional conformal field theories with Abelian symmetry,''
In *Shifman, M. (ed.) et al.: From fields to strings, vol. 2* 1173-1200.
[hep-th/0307041].
}

\lref\WittenABA{
  E.~Witten,
  ``Fermion Path Integrals And Topological Phases,''
[arXiv:1508.04715 [cond-mat.mes-hall]].
}

\lref\WuGE{
  T.~T.~Wu and C.~N.~Yang,
  ``Dirac Monopole Without Strings: Monopole Harmonics,''
Nucl.\ Phys.\ B {\bf 107}, 365 (1976).
}

\lref\XuNXA{
  F.~Xu,
  ``Algebraic coset conformal field theories,''
Commun.\ Math.\ Phys.\  {\bf 211}, 1 (2000).
[math/9810035].
}

\lref\XuLXA{
  C.~Xu and Y.~Z.~You,
  ``Self-dual Quantum Electrodynamics as Boundary State of the three dimensional Bosonic Topological Insulator,''
Phys.\ Rev.\ B {\bf 92}, no. 22, 220416 (2015).
[arXiv:1510.06032 [cond-mat.str-el]].
}

\lref\ZupnikRY{
   B.~M.~Zupnik and D.~G.~Pak,
   ``Topologically Massive Gauge Theories In Superspace,''
Sov.\ Phys.\ J.\  {\bf 31}, 962 (1988).
}

\lref\ZwiebelWA{
  B.~I.~Zwiebel,
  ``Charging the Superconformal Index,''
JHEP {\bf 1201}, 116 (2012).
[arXiv:1111.1773 [hep-th]].
}

\lref\Pufu{S.~Pufu, private communication and presentation at \hfill \break {\tt http://online.kitp.ucsb.edu/online/qft-c14/pufu} (2014)}


\lref\ABHS{``Enhanced Global Symmetries in $2+1$ Dimensions,'' to appear.}

%
%

\vskip-60pt
{\hfil SISSA 62/2016/FISI}
\Title{} {\vbox{\centerline{Chern-Simons-matter dualities}
\centerline{}
\centerline{with $SO$ and $USp$ gauge groups}
 }}

\vskip-15pt
\centerline{Ofer Aharony${}^1$, Francesco Benini${}^{2,3}$, Po-Shen Hsin${}^4$, and Nathan Seiberg${}^2$}
\vskip15pt
\centerline{\it ${}^1$ Department of Particle Physics and Astrophysics, Weizmann Institute of Science,}
\centerline{{\it Rehovot 7610001, Israel}}
\centerline{\it ${}^2$ School of Natural Sciences, Institute for Advanced Study, Princeton, NJ 08540, USA}
\centerline{\it ${}^3$ {SISSA, via Bonomea 265, 34136 Trieste, Italy �\char`\& \ INFN, Sezione di Trieste}}
\centerline{\it ${}^4$ Department of Physics, Princeton University, Princeton, NJ 08544, USA}
\vskip25pt

\noindent
In the last few years several dualities were found between the low-energy behaviors of Chern-Simons-matter theories with unitary
gauge groups coupled to scalars, and similar theories coupled to fermions. In this paper we generalize
those dualities to orthogonal and symplectic gauge groups. In particular, we conjecture dualities between $SO(N)_k$
Chern-Simons theories coupled to $N_f$ real scalars in the fundamental representation, and $SO(k)_{-N+N_f/2}$
theories coupled to $N_f$ real (Majorana) fermions in the fundamental.
For $N_f=0$ these are just level-rank dualities of pure Chern-Simons theories, whose precise form we clarify.
They lead us to propose new gapped boundary states of topological insulators and superconductors.
For $k=1$ we get an interesting low-energy duality between $N_f$ free Majorana fermions and an $SO(N)_1$ Chern-Simons theory coupled to $N_f$ scalar fields (with $N_f \leq N-2$).

\bigskip
\Date{November 2016}


%
%

\newsec{Introduction}

Three independent lines of investigation have recently converged on a long list of new dualities, which relate the low-energy (IR) behavior of two different $(2+1)d$ field theories.  One source of input came from the condensed matter literature (e.g.\ \refs{\PeskinKP\DasguptaZZ\BarkeshliIDA\SonXQA\PotterCDN-\WangGQJ}).  Another approach was based on the study of Chern-Simons (CS) theories coupled to matter in the fundamental representation with large $N$ and large $k$ with fixed $N/k$.  In some cases two different theories, one of them fermionic and the other bosonic, were argued \refs{\AharonyJZ,\GiombiKC,\AharonyNH} to be dual to the same Vasiliev high-spin gravity theory on $AdS_4$ (see e.g.\ \VasilievVF), and thus dual to each other by a duality exchanging strong and weak coupling.  Another approach to finding such dualities with finite $N$ and $k$ is based on starting with a pair of dual ${\cal N}=2$ supersymmetric theories \refs{\IntriligatorEX\deBoerMP\AharonyBX\AharonyGP\GiveonZN\KapustinGH\WillettGP\BeniniMF\AharonyCI
\newIS\AharonyDHA\ParkWTA-\AharonyKMA} and turning on a relevant operator that breaks supersymmetry.  If the flow to the IR is smooth, we should find a non-supersymmetric duality \refs{\JainGZA,\GurPCA}.  Motivated by this whole body of work, a number of dualities based on unitary gauge groups were conjectured in \AharonyMJS\ and elaborated in \HsinBLU:
\eqn\Oferdg{\eqalign{
N_f\ {\rm scalars\ with\ } SU(N)_k \quad &\longleftrightarrow \quad N_f\ {\rm  fermions\ with }\ U(k)_{-N+{N_f\over 2}, -N + {N_f\over 2}} \cr
N_f\ {\rm  scalars\ with\ } U(N)_{k,k}\quad &\longleftrightarrow \quad N_f\ {\rm  fermions\ with\ } SU(k)_{-N+{N_f\over 2}}\cr
N_f\ {\rm  scalars\ with\ }U(N)_{k,k\pm N}\quad &\longleftrightarrow \quad N_f\ {\rm  fermions\ with\ } U(k)_{-N+{N_f\over 2}, -N\mp k+{N_f\over 2}}
}}
for $N_f \leq N$. All CS theories here and below are viewed as the low-energy limits of the corresponding Yang-Mills-Chern-Simons theories; our conventions and more details on CS couplings and fermion determinants are collected in Appendix A. Here and throughout this paper we take $N, k \geq 0$; additional dualities follow from reversing the spacetime orientation in these dualities. The matter fields are in the fundamental representation of the gauge group and it is implicit that the scalars $\phi$ are at a $|\phi|^4$ fixed point. The $U(L)$ groups have two levels when $L>1$. See \HsinBLU\ for more details. The $N=k=N_f=1$ versions of these dualities were analyzed and coupled to appropriate background fields in \refs{\SeibergGMD,\KarchSXI,\MuruganZAL}, and their relation to supersymmetric dualities was analyzed in \refs{\KachruRUI,\KachruAON}, thus providing further evidence for their validity.

Our goal in this paper is to extend this line of investigation to orthogonal and symplectic gauge groups. We will conjecture the following IR dualities:%
\foot{For the symplectic groups our notation is $USp(2N)=Sp(N)$, and in particular $USp(2)=Sp(1)=SU(2)$. It is worth noting that for the orthogonal groups the level $k$, which is an integer, is normalized such that the Chern-Simons term is ${k\over 2\cdot 4\pi} \Tr \big( AdA + {2\over 3} A^3 \big)$ with a trace in the vector representation. As we will discuss below, the $SO(N)_k$ theory with even $k$ is a conventional non-spin topological quantum field theory (it does not require a spin structure), while the theory with odd $k$ is spin.}
\eqn\Oferdgo{\eqalign{
N_f\ {\rm real\ scalars\ with\ } SO(N)_k \quad &\longleftrightarrow \quad N_f\ {\rm real\ fermions\ with }\ SO(k)_{-N+{N_f\over 2}} \cr
N_f\ {\rm  scalars\ with\ } USp(2N)_k\quad &\longleftrightarrow \quad N_f\ {\rm fermions\ with\ } USp(2k)_{-N+{N_f\over 2}}
}}
The matter fields are in the fundamental representation of the gauge group.
The $SO$ dualities are conjectured to hold for $N_f \leq N-2$ if $k=1$, $N_f \leq N -1$ if $k=2$, and $N_f \leq N$ if $k>2$. Notice that these dualities, as opposed to the ones considered before, involve real scalars and real (Majorana) fermions. The $USp$ dualities are conjectured to hold for $N_f \leq N$.
In the 't~Hooft limit of large $N$ and $k$, with fixed $N_f$ and $N/k$, these dualities are supported by the same considerable evidence as their $U(N)$ counter-parts (since the orthogonal and symplectic theories are just projections of the $U(N)$ theories at leading order in $1/N$).

As a particularly interesting special case, if we set $k=1$ in the $SO$ dualities, we conjecture that the $SO(N)_1$ CS theory coupled to $N_f$ scalar fields in the vector representation (with any $N \geq N_f+2$) flows to $N_f$ free Majorana fermions.

In a companion paper \ABHS\ we will discuss a number of non-trivial fixed points with enhanced global symmetry.  Every one of them has a number of dual descriptions. The full global symmetry appears classically in some descriptions, but it only appears as a quantum enhanced low-energy symmetry in others. For example, taking special cases of \Oferdg, all the theories
\eqn\SpecialFourDual{\eqalign{
{\rm one\ scalar\ with}\ U(1)_2 \quad &\longleftrightarrow\quad {\rm one\ fermion\ with}\ U(1)_{-\frac32} \cr
\updownarrow \qquad\qquad\qquad&\qquad\qquad\qquad\quad \updownarrow \cr
{\rm one\ fermion\ with}\ SU(2)_{-\frac12} \quad &\qquad\quad\;\, {\rm one\ scalar\ with}\ SU(2)_1 \;
}}
lead to the same nontrivial fixed point with a global $SU(2) $ symmetry. The two descriptions at the bottom make the  $SU(2)$ global symmetry manifest at the classical level. (Actually, only $SO(3)$ acts faithfully on gauge invariant operators.) However, the symmetry only appears quantum mechanically in the two descriptions at the top.

The duality between the two theories at the bottom of \SpecialFourDual\ is the simplest example of the $USp$ dualities \Oferdgo, thus providing a nontrivial check of them. Also the duality between the two theories at the top of \SpecialFourDual\ appears among our $SO$ dualities for $SO(2)=U(1)$ \Oferdgo, but the $SO$ duality acts on the operators in a different way, that is related to the $U(1)$ duality by a global $SU(2)$ rotation. Thus, the enhanced global symmetry plays a crucial role in the consistency checks of the dualities \Oferdgo.

If we set $N_f=0$ in the suggested dualities \Oferdg\ and \Oferdgo, we find simple dualities involving topological quantum field theories (TQFTs):
\eqn\OferdgUO{\eqalign{
SU(N)_k \quad &\longleftrightarrow \quad  U(k)_{-N,-N} \cr
U(N)_{k,k\pm N}\quad &\longleftrightarrow \quad U(k)_{-N, -N\mp k} \cr
SO(N)_k \quad &\longleftrightarrow \quad SO(k)_{-N} \cr
USp(2N)_k\quad &\longleftrightarrow \quad  USp(2k)_{-N} \;.
}}
These are level-rank dualities in pure Chern-Simons theory. They provide one of the main motivations for the dualities \Oferdg, \Oferdgo, or conversely, they are a nontrivial check of them.  Although there exists a large literature about such level-rank dualities, we could not find a precise version of them. In \HsinBLU\ a careful analysis derived the first two lines in \OferdgUO\  and clarified that they hold, in general, only when the theories are spin-Chern-Simons theories (see \HsinBLU\ for details).  Below we will provide a similar proof of the orthogonal and symplectic dualities in \OferdgUO\ and will establish the need for the theories to be spin.%
\foot{However we will also find that for special values of $N,k$ the dualities \OferdgUO\ are valid for conventional TQFTs as well. In particular the first $SU/U$ duality is non-spin for $N$ even and $Nk=0\mod8$, the $SO$ duality is non-spin for $N,k$ even and $Nk=0\mod 16$, and the $USp$ duality is non-spin for $Nk=0\mod4$.}

The level-rank dualities can be used to show that several TQFTs, although not manifestly so, are time-reversal invariant at the quantum level. We provide a rich set of examples, summarized by the following table:

\bigskip\smallskip

\halign{
\hfil#\hfil&\hfil#\hfil&
\hfil#\hfil&
\hfil#\hfil&
\hfil#\hfil \cr
\noalign{\hrule\smallskip}
\omit\bf $\CT$-invariant theories \kern-5pt &  ${\bf N}$ & {\bf property} & {\bf framing anomaly} \cr
\noalign{\smallskip\hrule\smallskip}
$U(N)_{N,2N}$ & even $N$ & ${\cal T}$-invariant spin theory & $(N^2+1)/2$ \cr
 & odd $N$ & need to add $\psi$ & \cr
\noalign{\smallskip\hrule\smallskip}
$PSU(N)_N$ & even $N$ & ${\cal T}$-invariant spin theory & $(N^2-1)/2$ \cr
 & odd $N$ & ${\cal T}$-invariant non-spin theory & \cr
\noalign{\smallskip\hrule\smallskip}
$USp(2N)_N$ & even $N$ & ${\cal T}$-invariant non-spin theory & $N^2$ \cr
	& odd $N$ & need to add $\psi$ &  \cr
\noalign{\smallskip\hrule\smallskip}
$SO(N)_N$ & odd $N$ & ${\cal T}$-invariant spin theory & \cr
& $N=0$ mod 4 \kern10pt & ${\cal T}$-invariant non-spin theory & $N^2/4$\cr
& $N=2$ mod 4 \kern10pt & need to add $\psi$ & \cr
\noalign{\smallskip\hrule}
}

\bigskip\smallskip

\noindent
These theories represent new possible gapped boundary states of topological insulators and topological superconductors.

In Section 2 we analyze in detail the Chern-Simons-matter dualities for $USp(2N)$ gauge groups, and in Section 3 for $SO(N)$ groups. In Section 4 we comment on the relation to high-spin gravity theories on $AdS_4$. Section 5 gives a detailed description of level-rank dualities for orthogonal and symplectic groups, and Section 6 describes their implications for constructing new time-reversal-invariant TQFTs. Appendix A explains our notation and conventions.

After the completion of this work, we received \MetlitskiDHT\ where the duality of free fermions to $SO(N)_1$ with scalars is worked out.


\newsec{Dualities between $USp(2N)$ Chern-Simons-matter theories}

In Section 5 we will derive and discuss certain dualities of spin-TQFTs that take the form of level-rank dualities:
\eqn\spinLRdualUSpmain{
USp(2N)_k \times SO(0)_1 \quad\longleftrightarrow\quad USp(2k)_{-N} \times SO(4kN)_1 \;. }
In our conventions $USp(2N) = SU(2N) \cap Sp(2N,\C)$, and further details are collected in Appendix A.
We recall that a spin-TQFT---as opposed to a conventional topological quantum field theory---can only be defined on a manifold with a spin structure, and if multiple spin structures are possible, then the spin-TQFT will depend on the choice. A spin-TQFT always has a transparent line operator of spin $\frac12$. In \spinLRdualUSpmain\ the $USp$ factors are non-spin, while the $SO$ factors are trivial spin-TQFTs (discussed e.g.\ in \SeibergRSG), whose presence is important for the duality to work.

We can then add matter in the fundamental representation, bosonic on one side and fermionic on the other, and conjecture new boson/fermion dualities. This is done in such a way that renormalization group (RG) flows in which all matter becomes massive are consistent with \spinLRdualUSpmain. We thus propose the following dualities between the low-energy limits of Scalar and Fermionic theories:
\vskip 10pt
\centerline{\bf Theory S: A $USp(2N)_k$ theory coupled to $N_f$ scalars with $\phi^4$ interactions}
\vskip-13pt
\eqn\USpdualmatter{{\rm and}}
\vskip0pt
\centerline{\bf Theory F: A $USp(2k)_{-N + \frac{N_f}2}$ theory coupled to $N_f$ fermions}
\vskip 10pt
\noindent
for $N_f \leq N$. (The meaning of half-integer CS levels is the standard one, reviewed in Appendix A.)
More precisely, Theory S also includes the spin-topological sector $SO(0)_1$ that makes it into a spin theory and provides a transparent line of spin $\frac12$. Theory F is already manifestly spin, however to correctly reproduce its framing anomaly%
\foot{In general, the two sides of the duality have different framing anomalies (see \WittenHF). We fix that by adding a trivial spin-TQFT that has the difference in the anomaly. For our purposes this is the same as a gravitational Chern-Simons term with an appropriate coefficient.}
we include $SO\big( 4k(N- N_f)\big)_1$.
For completeness, let us rewrite the duality including its time-reversed version:
\eqn\USpdualmatterTrev{\eqalign{
USp(2N)_k \times SO(0)_1 {\rm \ with}\ N_f \ \phi_i \;&\leftrightarrow\; USp(2k)_{-N + \frac{N_f}2} \times SO\big( 4k (N - N_f) \big)_1 {\rm \ with}\ N_f \ \psi_i \cr
USp(2N)_{-k} \times SO(0)_1 {\rm \ with}\ N_f \ \phi_i \;&\leftrightarrow\; USp(2k)_{N - \frac{N_f}2} \times SO( 4kN)_{-1} {\rm \ with}\  N_f \ \psi_i}}
where $\phi_i$ are scalars and $\psi_i$ are fermions.

Theory S contains $N_f$ complex scalars in the fundamental $\bf{2N}$ representation of $USp(2N)$. Since the representation is pseudo-real, we can rewrite them in terms of $4NN_f$ complex scalars subject to the reality condition $\varphi_{ai} \Omega^{ab} \widetilde\Omega^{ij} = \varphi_{bj}^*$, where $a=1, \dots, 2N$ is a fundamental index of $USp(2N)$, $i=1,\dots, 2N_f$ is a fundamental index of $USp(2N_f)$, and $\Omega^{ab}$, $\widetilde\Omega^{ij}$ are the corresponding symplectic invariant tensors. This description makes the $USp(2N_f)$ flavor symmetry of the theory manifest. There is one quadratic term and two%
\foot{For $N_f=1$ there is only one independent quartic term.}
quartic terms one can write that preserve the $USp(2N_f)$ flavor symmetry: in terms of the antisymmetric meson matrix $M_{ij} = \varphi_{ai} \Omega^{ab} \varphi_{bj}$ they are
\eqn\USpDefTerms{
\CO^{(2)} = \Tr( \widetilde\Omega M) = \sum\nolimits_{ai} |\varphi_{ai}|^2 \;,\qquad \CO^{(4)}_1 = \big(\! \Tr (\widetilde\Omega M) \big)^2 \;,\qquad \CO^{(4)}_2 = \Tr (\widetilde\Omega M \widetilde\Omega M) \;. }
All these three terms are relevant in a high-energy Yang-Mills-Chern-Simons theory with these matter fields.  We turn them on with generic coefficients and make a single fine-tuning (which we can interpret as the coefficient of the quadratic term).  We conjecture that by doing that the long distance theory is at an isolated nontrivial fixed point.  It is a generalization of the Wilson-Fisher fixed point, and it has a single $USp(2N_f)$-invariant relevant deformation which may be identified with $\CO^{(2)}$. Theory S is defined at such a fixed point.

Note that the $\Z_2$ center of the $USp(2N)$ gauge group acts in exactly the same way as the $\Z_2$ center of the $USp(2N_f)$ global symmetry group. So local gauge-invariant operators actually sit in representations of $USp(2N_f)/\Z_2$ (e.g. for $N_f=1$ they always have integer spin under $SU(2)$). However we can have non-local operators like a $\varphi_{ai}$ attached to a Wilson line, where the charge of the Wilson line under the center of $USp(2N)$ is correlated with the charge of the operator that it ends on under the center of $USp(2N_f)$. In this sense the gauge $\times$ global symmetry group is $\big( USp(2N) \times USp(2N_f) \big)/\Z_2$.

Theory F contains $N_f$ complex spinors in the fundamental $\bf{2k}$ representation of $USp(2k)$, and again, rewriting them in terms of $4kN_f$ Dirac spinors with a reality condition $\psi_{ai} \Omega^{ab} \widetilde\Omega^{ij} = \psi^c_{bj}$ (where $\psi^c$ is the charge conjugate) makes the $USp(2N_f)$ flavor symmetry manifest. At high energies the corresponding Yang-Mills-Chern-Simons theory has a single relevant $USp(2N_f)$-invariant operator, which is the quadratic mass term. We tune it to zero in the IR, and assume it is the only $USp(2N_f)$-invariant relevant operator there. The bare CS levels (see Appendix A) are $(N_f-N)$ in \USpdualmatter\ and $N$ in \USpdualmatterTrev.

Notice that while the spin-TQFTs involved in the level-rank duality \spinLRdualUSpmain\ have a $\Z_2$ one-form global symmetry associated to the center of the group, such a symmetry is broken in the theories with matter because the latter transforms under the center \GaiottoKFA. In fact, the theories in \USpdualmatter\ do not have any discrete global symmetries.

\subsec{RG flows}

We cannot prove the dualities in \USpdualmatter, however we can perform some consistency checks. For instance, we can connect different dual pairs by RG flows triggered by mass deformations. In both Theories S and F, turning on a mass at high energies leads to turning on the unique relevant deformation of the low-energy conformal field theory (CFT). So turning on a bosonic mass-squared $m_\phi^2$ in Theory S should have the same effect at low energies as turning on a fermion mass $m_\psi$ in Theory F.

In Theory S, if we give a positive mass-squared to one of the complex scalars we simply reduce $N_f$ by one unit. However, if we turn on a negative mass-squared, a complex scalar condenses Higgsing the gauge group to $USp(2N-2)_k$, in addition to reducing the number of flavors. In Theory F, when giving mass to one of the complex fermions, the phase of its partition function becomes either $e^{-i\pi\eta(A)}$ or $1$ in the IR limit, depending on the sign of the mass. In both cases the number of flavors is reduced by one, however in the first case one can use the APS index theorem \AtiyahJF\ (see Appendix A) to rewrite the leftover regularization term $e^{-i\pi\eta(A)}$ in terms of a shift of the bare gauge and gravitational CS terms. Thus, tuning the mass of the remaining $(N_f-1)$ flavors to zero, the RG flow leads\foot{Here we assume that the RG flow still leads to a non-trivial CFT with a single $USp(2N_f-2)$-invariant relevant operator.} to the following pairs:
\eqn\USpflows{\eqalign{
& USp(2N)_k \times SO(0)_1 {\rm \ with}\ \phi_i \quad\leftrightarrow\quad USp(2k)_{-N + \frac{N_f-1}2} \times SO\big( 4k(N-N_f+1) \big)_1 {\rm \ with}\ \psi_i \cr
& USp(2N-2)_k \times SO(0)_1 {\rm with}\ \phi_i \quad\leftrightarrow\quad USp(2k)_{-N + \frac{N_f+1}2} \times SO\big( 4k (N- N_f) \big)_1 {\rm \ with}\ \psi_i
}}
each with $N_f-1$ matter fields. These are consistent with the proposed duality.

If $N_f < N$, by flowing with different combinations of positive and negative masses squared, in Theory S the gauge group can range between $USp(2N)_k$ and $USp\big(2(N-N_f)\big)_k$, while in Theory F the level can range between $USp(2k)_{-N}$ and $USp(2k)_{-N+N_f}$. In all these cases we do not find any inconsistency in the duality. Starting with a dual pair with $N_f = N$, we can give negative mass-squared to all flavors and flow to a dual pair with $N_f = N = 0$. In this case, Theory S is gapped and empty in the IR; Theory F is a $USp(2k)_0$ Yang-Mills theory, which confines and has a single gapped ground state. Thus the duality is still valid.

On the other hand, consider the case $N_f \ge N+1$ and turn on generic masses for all the flavors.  Theory F flows to a non-trivial topological theory.  Let us compare with Theory S.  Here generic negative mass-squared for all matter fields Higgses the gauge group completely.  The IR theory could be gapped or could have massless Goldstone bosons, but since the gauge group is completely Higgsed, it cannot include a topological sector.  Hence, the duality cannot be correct in this case. We conclude that none of the pairs with $N_f \geq N+1$ can be dual.

\subsec{Coupling to background gauge fields}

Given that our system has a global $USp(2N_f)$ symmetry, we can couple it to background gauge fields for that symmetry.  Our goal here is to identify the CS counterterms \ClossetVP\ for these fields that are needed for the duality.

Let  us start with the scalar side of the duality.  We start with a $USp(2N)_k$ CS theory for the dynamical fields and we can also have $USp(2N_f)_{k_s}$ for some integer $k_s$ for the classical fields. If we give masses to all scalars such that the gauge symmetry is not Higgsed, then the low energy theory is purely topological.  It is a $USp(2N)_{k}\times USp(2N_f)_{k_s} $ CS theory, where the second factor is classical.  In the fermionic side of the duality we start with $USp(2k)_{-N+{N_f\over 2}}$ for the dynamical fields and $USp(2N_f)_{k_f}$ for the classical fields.  These mean that the bare CS levels for these two groups are $-N+N_f$ and $k_f +{k\over 2}$ respectively.  Repeating the mass deformation of the bosonic side we find at low energy a topological $USp(2k)_{-N}$ as well as a CS counterterm for the classical fields $USp(2N_f)_{k_f-{k\over 2}}$.  For this to match with the bosonic side we must choose%
\foot{It is common in the literature to argue for 't Hooft-like anomaly matching conditions restricting the level of CS terms for the global symmetry. These levels could be half-integral in theories with fermions and they are always integral in theories with bosons. Based on that, one might attempt to exclude many of these boson/fermion dualities.  Instead, as in Appendix A, the fractional part of the CS terms always arise from the dynamics and the bare CS levels are always integral. In the cases where our theories have fermions leading to a fractional level, the bosonic dynamics in the other side of the duality must lead to equal fractional levels.  Indeed, our analysis of the renormalization group flow out of the fixed point is consistent with this assertion. However, as we will now show, there do exist non-trivial 't Hooft-like anomaly matching of the integer bare CS terms for the global symmetries.}
\eqn\kNNfmat{
k_f=k_s+{k\over 2} \;.
}
Of course, we have the freedom to add the same CS counterterm on the two sides of the duality. This will add an arbitrary integer to $k_s$ and the same integer to $k_f$.

We can repeat the same considerations with an opposite sign for the mass deformations. In the scalar theory, Higgsing occurs and the symmetry group is reduced to $USp\big(2 (N-N_f)\big)_k \times USp(2N_f)_{k_s+k}$, where the broken part of the gauge group is identified with the flavor group and this causes the shift of the CS counterterm for the classical fields. In the fermionic theory we find $USp(2k)_{-N+N_f} \times USp(2N_f)_{k_f+\frac k2}$. As a non-trivial check, equality on the two sides requires the very same relation \kNNfmat.

As we discussed above, the global symmetry that acts faithfully on local operators is $ USp(2N_f)/\Z_2$ and this puts restrictions on the CS counterterms.  More precisely, in the bosonic side we would like the bare CS terms to be consistent for $\big( USp(2N)_k\times USp(2N_f)_{k_s} \big)/\Z_2$ and the $\Z_2$ quotient is consistent only for
\eqn\kNNfco{Nk +N_f k_s \;\in\; 2\Z \;.}
In the fermionic side the bare CS terms are  $\big( USp(2k)_{-N+N_f}\times USp(2N_f)_{k_s +k} \big)/\Z_2$, where we used \kNNfmat.  This is consistent for
\eqn\kNNfcoco{-kN +2kN_f +k_sN_f \;\in\; 2\Z \;.}
Fortunately, \kNNfco\ is the same condition as \kNNfcoco.  In the spirit of 't~Hooft anomaly matching, this is a non-trivial consistency check on our duality.  The obstruction to the $\Z_2$ quotient is the same in the two sides of the duality.

When the condition \kNNfco\ is not satisfied, we cannot mod out by the $\Z_2$ and fewer backgrounds of the gauge fields are allowed.  In those cases it might still be possible to extend the $USp(2N_f)$ classical gauge fields to a $(3+1)d$ bulk and consistently take the $\Z_2$ quotient there.

\subsec{Small values of the parameters}

It is instructive to look at the dualities \USpdualmatter\ for small values of $N$, $k$ and $N_f$. We already discussed that the case $N=N_f=0$ is the statement that $USp(2k)_0$ confines with a single vacuum. The case $k=0$ is the statement that $USp(2N)_0$ with $N_f \leq N$ complex scalars confines with a single vacuum: although we have no proof that this is true, it is surely plausible.

As discussed around \SpecialFourDual, the case $N= k = N_f=1$ can be derived from the dualities in \refs{\AharonyMJS,\HsinBLU}, giving us more confidence that the duality is correct.

\subsec{New fermion/fermion and boson/boson dualities}

Combining the dualities for symplectic groups in \USpdualmatter\ with those for (special) unitary groups in \refs{\AharonyMJS,\HsinBLU}, we can find new fermion/fermion and boson/boson dualities.

For instance, we can take \USpdualmatter\ with $N = N_f=1$ and combine it with the first duality of (5.5) in \HsinBLU. This gives us a fermion/fermion duality
\eqn\USpFermFerm{
USp(2k)_{-\frac12} {\rm \ with\ 1\ fermion\ in\ } {\bf{2k}} \qquad\longleftrightarrow\qquad U(k)_{-\frac32} {\rm \ with\ 1\ fermion\ in\ } \bf{k} \;.}
(To be precise, the theory on the right should include a decoupled $SO(2k)_1$ factor.) As we discussed, the duality with $k=1$ can be derived from \refs{\AharonyMJS,\HsinBLU}, but the other ones are new.

Note that for all $k$ the theories in \USpFermFerm\ have a global $SU(2)$ symmetry, which is manifest in the LHS of \USpFermFerm.  On the RHS of the duality we have a manifest $U(1)$ monopole number symmetry and charge conjugation $\CC$, which does not commute with it because it maps the monopole number $n$ to $-n$.  Our duality suggests that this $U(1) \rtimes \Z_2^\CC$ classical symmetry is enhanced in the quantum theory to $SU(2)$. The currents that extend the Abelian symmetry to $SU(2)$ must carry monopole charge. We suggest that they are constructed out of the monopole operator and its conjugate in the $U(k)$ theory.  Since these carry charge, each of them should be dressed by a fermion to be gauge-invariant.

Similarly, from \USpdualmatter\ with $k=1$ and \refs{\AharonyMJS,\HsinBLU} we can obtain the boson/boson duality
\eqn\USpBosBos{
USp(2N)_1 {\rm \ with}\ N_f {\rm \ scalars \ in}\ {\bf{2N}} \qquad\longleftrightarrow\qquad U(N)_2 {\rm \ with}\ N_f {\rm \ scalars\ in}\ \bf{N} }
with $N_f \leq N$.
Both sides should include $SO(0)_1$ and be regarded as spin theories. The case $N=N_f=1$ was already found in \refs{\AharonyMJS,\HsinBLU}, but the other ones are new.  As above, the global symmetry of these theories is $USp(2N_f)$, which is manifest in the LHS, thus we conjecture that the manifest $U(N_f)$ and charge conjugation symmetries in the RHS are enhanced to $USp(2N_f)$.

\newsec{Dualities between $SO(N)$ Chern-Simons-matter theories}

For orthogonal groups there exists a similar level-rank duality of spin-TQFTs (derived in Section 5):
\eqn\spinLRdualSOmain{
SO(N)_k \times SO(0)_1 \quad\longleftrightarrow\quad SO(k)_{-N} \times SO(kN)_1 \;. }
(For our conventions on CS terms see Appendix A and footnote 1.) Arguments similar to those of the previous section, and to those used for $SU(N)$ and $U(N)$ groups, suggest a duality between the low-energy limits of:
\vskip 10pt
\centerline{\bf Theory S: An $SO(N)_k$ theory coupled to $N_f$ real scalars with $\phi^4$ interactions}
\vskip-13pt
\eqn\SOdualmatter{{\rm and}}
\vskip0pt
\centerline{\bf Theory F: An $SO(k)_{-N + \frac{N_f}2}$ theory coupled to $N_f$ real fermions.}
\vskip 10pt
\noindent
As we explain below, the duality can only be true for $N_f \leq N-2$ if $k=1$, for $N_f \leq N-1$ if $k=2$, and for $N_f \leq N$ if $k>2$. Moreover, one should remember that Theory S includes the trivial spin-topological sector $SO(0)_1$, while Theory F includes $SO\big( k (N- N_f) \big)_1$. The matter fields are all in the vector representation.
There is considerable evidence for this duality at large $N$ and $k$; at finite values of $N$ and $k$ we can check its consistency by mass flows, including flows to the level-rank dualities between pure Chern-Simons theories \spinLRdualSOmain\ described in Section 5.

The two low-energy theories are defined by tuning the masses to zero, and assuming that both sides flow to a fixed point, which has a single relevant operator consistent with the global symmetries of the high-energy theory.
For finite values of $N$, $k$ and $N_f$ we do not know when this is true; additional operators could become relevant at low energies, which would prevent the two theories from flowing to the same fixed point, or the fixed point may cease to exist (say because a mass gap develops, or because we spontaneously break some of the flavor symmetries on one side but not the other). The limits on $N_f$ above arise because we show that for larger values of $N_f$ the duality cannot hold, but we conjecture that it does hold for the values mentioned above.

\subsec{Flows}

We can flow from the duality with $(N,k,N_f)$ to the dualities with $(N,k,N_f-1)$ and with $(N-1,k,N_f-1)$ by adding a mass for a single flavor. Both the mass-squared deformation in Theory S, and the mass deformation in Theory F, are expected to flow to the same relevant operator in the low-energy CFT (which is in the symmetric tensor representation of the $SO(N_f)$ flavor group). Previously we mentioned the $SO(N_f)$-invariant component of this relevant operator, but we assume here that it also has a symmetric-traceless counter-part that is relevant.
In theory F we can integrate out the massive flavor, schematically shifting the Chern-Simons level of the remaining low-energy theory by $\pm {1\over 2}$, depending on the sign of the mass. In theory S the same flavor, depending on the sign of its mass-squared, either becomes massive, or condenses and breaks the gauge group to $SO(N-1)$ (this is all similar to the $U$ and $USp$ cases). In this flow we fine-tune the mass of the remaining $(N_f-1)$ flavors to zero. The flow to the lower duality exists whenever no additional operator becomes relevant, and whenever the new IR CFT exists; under these assumptions it leads to a new duality between the lower theories S$'$ and F$'$.

Reversing this logic, if we have a non-trivial duality for some values of $(N,k,N_f)$, we can assume that the duality holds for all $(N',k,N_f')$ with 
$N + N_f' - N_f \geq N' \geq N$ and $N_f' > N_f$. If for some such value the duality fails but we still have a non-trivial theory with no extra relevant operators, then we expect the duality to fail also for the corresponding higher values (since otherwise we get a contradiction by first flowing to the ``higher'' IR CFT and then performing the mass flow).

When giving a mass to all $N_f$ fermions, Theory F flows to a pure $SO(k)$ Chern-Simons theory with a level between $(-N)$ and $(N_f-N)$, depending on the signs of the masses of the different flavors. Depending on the same signs, some flavors could condense in theory S, so that its gauge group is between $SO(N)_k$ and $SO(N-N_f)_k$. The resulting theories are then dual by level-rank dualities of $SO(N)$ spin-TQFTs (see Section 5), giving a consistency check on our dualities. Note that this assumes $N_f < N$.

For $N_f=N$ and an appropriate sign of the mass deformation, the gauge theory in Theory S is completely broken, and for generic values of the masses all scalars become massive and the theory develops a mass gap. In Theory F for the same masses the low-energy Chern-Simons level vanishes and the fermions are massive, so again we expect a mass gap.
For $N_f>N$ with the same choice of signs for the masses for all the flavors, we break the gauge group completely in Theory S, and generically all scalars are massive and we have a mass gap with a trivial theory at low energies. On the other hand, in Theory F for the same choice, all fermions become massive, but we end up with a non-trivial topological theory, so the duality necessarily breaks down for this case, as in the $U(N)$ and $USp(2N)$ cases.

\subsec{Global symmetries}

The UV Yang-Mills-Chern-Simons theories we start from have an $O(N_f)$ flavor symmetry, as well as two discrete symmetries discussed below.  The definition of the IR Theories S and F involves flows from these UV theories that preserve these $SO(N_f)$ and discrete symmetries. The $SO(N_f)$ symmetry allows for a single mass term which needs to be tuned, both on the scalar and on the fermion sides. As in \USpDefTerms, for $N_f>1$ the UV description of Theory S has two possible $\phi^4$ terms.  In terms of $M_{ij}=\phi_{ai}\phi_{aj}$ ($i,j$ are flavor indices and $a$ is a color index) they are $\CO^{(4)}_1 = \big(\! \Tr (M) \big)^2$ and $\CO^{(4)}_2 = \Tr (M^2)$ . We assume that for generic couplings of these operators they do not lead to any new relevant operator at low energies.%
\foot{For some values of $N$, $k$ and $N_f$ there are additional fixed points where one or both quartic operators are tuned to zero, and that also have fermionic Gross-Neveu-like duals, but we will not discuss them here.}
For some small values of $N$ and $k$ there are enhanced continuous symmetries, which we will discuss below.

In addition there are two discrete symmetries, that were discussed in detail in \AharonyKMA. There is a global ``charge conjugation'' symmetry $\CC$, which acts on the matter fields as $\phi_{1i} \to -\phi_{1i}$ and all other $\phi_{ai}$ are invariant. When $\CC$ is gauged the gauge group changes from $SO(N)$ to $O(N)$.%
\foot{For $N$ even, this symmetry always exchanges the two spinor representations of $Spin(N)$. They are in fact complex conjugate representations for $N=2\mod 4$, but not for $N=0\mod 4$.}  The $SO(N)$ vector indices may be contracted to form singlets either with $\delta_{ab}$ or with $\epsilon_{a_1 a_2 \cdots a_N}$. Operators that involve the latter contraction are odd under $\CC$, while all others are even. Since the product of two epsilon symbols may be replaced by a sum of products of $\delta$'s, the symmetry is $\Z_2$.

In the fermionic $SO(k)$ theory, the lowest-dimension operator charged under $\CC$ is a baryon operator, involving $k$ fermions contracted with an epsilon symbol. Classically the dimension of this operator is $k$, and in the quantum theory it has some anomalous dimension. Its Lorentz$\,\times SO(N_f)$ representation is a symmetric product of $k$ spinors which are vectors of $SO(N_f)$. In the scalar $SO(N)$ theory a similar contraction vanishes (for $N_f < N$) because of the statistics of the scalar operators. For $N_f=1$ the lowest-dimension $\CC$-odd operator comes from choosing $N$ different derivative operators acting on the scalar, and then contracting them with an epsilon symbol. Similarly for $N_f > 1$ we need to choose $N$ different combinations of derivatives and flavor indices; when $N_f=N$ we can contract $N_f$ different scalars with no derivatives. In the large $N$ limit with fixed $N_f$, the classical dimension of the lightest $\CC$-odd operator scales as $N^{3/2}$ \ShenkerZF. The Lorentz representation of these operators comes from the product of those of the derivative operators that we need to use.

Monopole operators in an $SO(N)$ theory are characterized by having some quantized flux around them, which can be chosen to be in the Cartan algebra of $SO(N)$ (this is a semi-classical characterization; in the quantum theory operators with different GNO charges can mix). The lowest one carries one unit of flux under a single $SO(2)$ subgroup of $SO(N)$. The flux breaks the $SO(N)$ gauge group to $\big(O(2)\times O(N-2)\big) / \Z_2$. The smallest monopole charge (which we normalize to be $1$) in the $SO(N)$ theory is defined by requiring mutual locality with matter fields (or Wilson lines) in the vector representation of $SO(N)$. In a $Spin(N)$ theory we need to require mutual locality also with fields (or Wilson lines) in the spinor representation, and thus the minimal monopole charge is $2$. This implies that the monopoles carry a $\Z_2$ global ``magnetic'' symmetry $\CM$, and operators that are allowed in $SO(N)$ but not in $Spin(N)$ are odd under $\CM$.
Note that unlike in $U(N)$ theories, the monopoles do not carry a $U(1)$ global charge (except when the gauge group is $SO(2)$), but a $\Z_2$ charge.

This description of the monopole operator is not manifestly $SO(N)$ invariant.  Alternatively, we can define this operator by removing a point from our spacetime and specifying a non-trivial bundle on the sphere that surrounds it.  Specifically, this monopole corresponds to having nontrivial second Stiefel-Whitney class $w_2$ on that sphere.  This makes it clear that the $\CM$ charge is a $\Z_2$ charge.

Gauge-invariance requires that a monopole in the $SO(N)_k$ Theory S must come together with $k$ fields charged under the $SO(2) \subset SO(N)$ gauge group. Charged scalars in the monopole background carry spin $\frac12$ \WuGE, and their scaling dimension is shifted from $\frac12$ to $1$. Thus the lightest monopole operator that is charged under $\CM$ has classical dimension $k$, and lies in a Lorentz$\,\times SO(N_f)$ representation that is a symmetric product of $k$ spinors that are fundamentals of $SO(N_f)$.

Charged fermions in a monopole background have integer spins, and in particular each fermion has a zero mode, and defining the monopole operator requires quantizing these $N_f$ zero modes. After quantizing the zero modes, one has to add in the $SO(k)_{-N+\frac{N_f}2}$ Theory F $N$ additional fermionic operators charged under $SO(2)$, of various integer spins, in order to form a gauge-singlet.

The Lorentz$\,\times SO(N_f)$ representations of these fermionic monopole operators are identical to those of the baryons in Theory S that were described above, and their classical dimensions are also the same (since the fermion modes in the monopole background have the same Lorentz quantum numbers and dimensions as derivatives acting on scalars). In particular their classical dimension in the large $N$ limit scales as $N^{3/2}$ \Pufu\ (and this statement is true also quantum mechanically \RadicevicYLA). Similarly, the lightest baryon operator in the fermionic theory has precisely the same quantum numbers and classical dimension as the lightest monopole operator in the scalar theory.

Above we were not careful about precisely which monopole we choose in the $SO(2)$ theory, and how it transforms under charge conjugation; these issues were discussed in detail in \AharonyKMA, and we review the discussion here. In an $SO(2)  =  U(1)$ theory, there are monopoles $V_n$ that carry $n$ units of the $U(1)_J$ magnetic charge (topological charge). Charge conjugation in this theory takes $n$ to $(-n)$, so we have one lightest monopole $(V_1+V_{-1}) = V_+$ which is $\CC$-even, and another $(V_1-V_{-1})=V_-$ which is $\CC$-odd. We can choose the $SO(N)$ monopole discussed above to correspond to either one of these operators in the $SO(2)$ subgroup. However, since the precise gauge group that remains in the monopole background is $\big( O(2)\times O(N-2) \big)/\Z_2$, if we choose the $\CC$-even monopole operator in $SO(2)$, we need to dress it by a $\CC$-even operator in $SO(N-2)$, while if we choose the $\CC$-odd monopole in $SO(2)$, it has to come with a $\CC$-odd operator in $SO(N-2)$, in order to be $SO(N)$-gauge-invariant. The monopoles we discussed above in $SO(N)$ theories were singlets of $SO(N-2)$, so they are $\CC$-even and involve the monopole $V_+$ in the $O(2)$ subgroup. In order to form a $\CC$-odd monopole operator (which was called a monopole-baryon in \AharonyKMA) we need to take $V_-$ and multiply it  by a $\CC$-odd operator in $O(N-2)$, namely a product of $(N-2)$ matter fields contracted with an epsilon symbol (in addition to the extra fields required for canceling the $SO(2)$ charge of the monopole). Repeating the same arguments as above, we find that the lightest monopole-baryon-operator in both theories F and S has a classical dimension scaling as $N^{3/2}$ in the large $N,k$ limit with fixed $N_f$, and the operators also lie in identical Lorentz$\,\times SO(N_f)$ representations in the two theories.

The arguments above strongly suggest that the duality exchanges monopoles and baryons, and takes the monopole-baryon operators to themselves, namely it exchanges the two $\Z_2$ global symmetries $\CC$ and $\MM$. In fact, we can see that this must be the case by performing the mass flow to the pure Chern-Simons theories, and noting (see Section 5) that the level-rank duality in these theories indeed exchanges $\CC$ with $\MM$. So this gives a nice consistency check for the duality. The fact that the classical dimensions on both sides match (at least at large $N$) is somewhat surprising, since one would expect their dimension to receive quantum corrections (except for the monopole operator in the fermionic theory which was shown in \RadicevicYLA\ to receive no quantum corrections to its dimension in the 't Hooft large $N$ limit). This is all very similar to the duality between $SU(N)$ and $U(k)$ CS-matter theories, which also exchanges baryon number with monopole number \refs{\RadicevicYLA,\AharonyMJS}.

By gauging $\CC$ and/or $\CM$ we can find related dualities involving $O(N)$, $Spin(N)$ and $Pin^{\pm}(N)$ gauge theories.  These theories can have additional labels, which are the coefficients of terms like $w_2w_1$ of the gauge bundle \AharonyKMA. These can be thought of as CS terms of various discrete gauge fields or as discrete theta parameters analogous to those studied in \refs{\AST,\KapustinGUA,\GaiottoKFA}. We will not discuss them here.

In the ${\cal N}=2$ supersymmetric version of the Chern-Simons-matter dualities between $SO(N)$ gauge theories \AharonyKMA, the duality maps $\CC_{\rm SUSY}$ to itself, while mapping $\CM_{\rm SUSY}$ to $\CC_{\rm SUSY} \CM_{\rm SUSY}$. Given the fact that one can flow from the supersymmetric theories to the pure Chern-Simons theories, and perhaps also to the non-supersymmetric Chern-Simons-matter theories along the lines of \refs{\JainGZA,\GurPCA}, this is confusing. The resolution is that in the ${\cal N}=2$ supersymmetric Chern-Simons-matter theories there is also a complex gaugino field in the adjoint representation. It has $(N-2)$ zero modes in the monopole background, which are a vector of $SO(N-2)$, and their product is odd under $\CC_{\rm SUSY}$. Thus the minimal monopole in the supersymmetric theory, that carries those zero modes, has an opposite charge-conjugation transformation from the minimal monopole in the non-supersymmetric theory (bosonic or fermionic). So we have $\CM_{\rm SUSY} = \CM$, but $\CC_{\rm SUSY} = \CC \CM$ (where $\CC$ and $\CM$ are defined as above in the non-supersymmetric theory). With this relation, the two dualities are consistent with the flows and identifications discussed above.

\subsec{Small values of $N$ and $k$}

When the theories on both sides are non-Abelian it is difficult to check the dualities. However, for $k = 1,2$, and for $N_f=N-1,N$, we can (possibly after Higgsing) have Abelian or empty gauge groups, so we can test the dualities in more detail.

When $k=1$ or $N=1$ we have no gauge group on one of the two sides, and also no magnetic symmetry, though the charge conjugation symmetry remains and changes the sign of the matter fields. On the fermionic side for $k=1$ we have $N_f$ free real (Majorana) fermions. On the bosonic side for $N=1$ we have an $O(N_f)$-invariant Wilson-Fisher fixed point, which arises by turning on the quartic operator in the theory of $N_f$ real scalars.

When $k=2$ or $N=2$ we have an Abelian gauge theory on one of the two sides. In this theory the magnetic $\Z_2$ symmetry $\CM$ is enhanced to a $U(1)_J$ symmetry (equal to $\CM$ modulo 2). As mentioned above, the charge conjugation symmetry does not commute with $U(1)_J$: it takes a monopole of $U(1)_J$ charge $n$ to one of charge $(-n)$. In addition, the UV theory in this case is equal to a $U(1)$ Maxwell-Chern-Simons theory (since $U(1)_{k'}$ with $N_f$ flavors is identical to $SO(2)_{k'}$ with $N_f$ flavors) which has, for $N_f > 1$, an enhanced $SU(N_f)$ flavor symmetry.

In the bosonic $N=2$ theory, the mass operator (which we can write in the $U(1)$ language as ${\cal O}^{(2)} = \Phi_i^{\dagger} \Phi_i$ in terms of complex fields $\Phi_i$ carrying $U(1)$ charge $1$) is $SU(N_f)$-invariant. For $N_f=2$, we have additional gauge-invariant quadratic operators: there is an operator charged under the flavor $SO(2)$ containing $|\Phi_1|^2 - |\Phi_2|^2$ and $(\Phi_1^{\dagger} \Phi_2 + \Phi_2^{\dagger} \Phi_1)$ which is $\CC$-even, and a flavor $SO(2)$ singlet ${\cal O}^{(2)}_2 = (\Phi_1^{\dagger} \Phi_2 - \Phi_2^{\dagger} \Phi_1)$ which is $\CC$-odd. So in flows from the UV Yang-Mills-Chern-Simons theory that preserve both $SO(N_f)$ and $\CC$ we do not need to consider either one of them, and we still require a single fine-tuning at low energies. For $N_f>2$ there is only one quadratic $SO(N_f)$-invariant operator.

At quartic order there is an $SU(N_f)$-invariant $\CC$-even operator $({\cal O}^{(2)})^2 = \Phi_i^{\dagger} \Phi_j^{\dagger} \Phi_i \Phi_j$. For $N_f > 1$, though, there is another $SO(N_f)$-invariant $\CC$-even quartic operator $\Phi_i^{\dagger} \Phi_i^{\dagger} \Phi_j \Phi_j$. When we view the theory as an $SO(2)$ gauge theory, and in particular when we flow to it from higher $SO(N)$ gauge theories, we only preserve the $SO(N_f)$ symmetry, so the latter operator is also turned on during the flow. We expect this extra operator to be irrelevant at the $SU(N_f)$-invariant fixed point of the $U(1)$ theory, and if so then the $SO(2)$ flow also reaches the same fixed point, at least for some values of the deformations from the UV theory. However, we do not know how to prove that this is the case, and there may be a different flow if the $SU(N_f)$-non-invariant operator is turned on with a large coefficient. We will assume below that we do end up at the same fixed point. For $N_f=2$ this extra operator is a linear combination of $({\cal O}^{(2)})^2$ and $({\cal O}^{(2)}_2)^2$, and there is another $SO(2)$-invariant $\CC$-odd operator ${\cal O}^{(2)} {\cal O}^{(2)}_2$ which will not be turned on in a $\CC$-invariant flow (an additional operator coming from the $SO(2)$-charged operator squared is a linear combination of these).

In the fermionic $k=2$ theory we have similar quadratic operators, and the quartic operators are irrelevant in the UV.

Thus, in both theories we expect the $SO(N_f)$-invariant $\CC$-invariant flow of the $SO(2)$ theory to end up at the same fixed point as that of the $SU(N_f)$-invariant $U(1)$ flow. We can then use known facts about the latter flow to learn about the status and implications of the $SO(N)$ dualities.

Now let us discuss the dualities for various small values of $k$ and $N$.

\subsec{The $k=1$ case}

For $k=1$ we have a duality between the theory of $N_f$ free real (Majorana) fermions, and an $SO(N)_1$ theory coupled to $N_f$ real Wilson-Fisher scalars.

For $N=1$ (implying also $N_f=1$) this duality is obviously wrong, since the usual Wilson-Fisher fixed point is not free (and thus the duality cannot hold also for higher $N=N_f$ which can flow to this value).

For $N=2$ we have on the bosonic side the fixed point of $U(1)_1$ coupled to Wilson-Fisher scalars. In this case we need $N_f=1$, and then the $U(L)$ dualities \refs{\AharonyMJS,\SeibergGMD}, assuming they hold for $L=1$, imply that this scalar theory is dual to a free complex fermion. So under this assumption the $SO$ duality cannot be correct for $(N,k,N_f) = (2,1,1)$, and thus also for $k=1$ and higher values of $N$ with $N_f=N-1$.

For higher values of $N$, with $N_f \leq N-2$, we cannot rule out the $k=1$ duality by known results. Thus we conjecture an IR duality between:
\vskip 10pt
\centerline{\bf Theory S: An $SO(N)_1$ theory coupled to $N_f$ real scalars with $\phi^4$ interactions}
\vskip-13pt
\eqn\BosonizedMajorana{{\rm and}}
\vskip0pt
\centerline{\bf Theory F: $N_f$ free Majorana fermions.}
\vskip 10pt
\noindent
As discussed above, the monopole operator of the scalar theory maps to the real fermion. The lowest case is $N=3$ and $N_f=1$, where we have an $SO(3)_1$ theory with a single scalar in the vector (adjoint) representation flowing to a free Majorana fermion. Since Theory F in this case has no magnetic symmetry, the duality implies that all baryons and monopole-baryons of Theory S decouple at low energies.

\subsec{The $N=1$ case}

In this case, since we should have $N_f=1$, the scalar theory is just a real Wilson-Fisher scalar. The dual fermionic theory has a real fermion coupled to an $SO(k)_{-\frac12}$ CS theory. The case $k=1$ was already ruled out above. The case $k=2$ is related to a $U(1)_{-\frac12}$ theory, which maps by the dualities of \refs{\AharonyMJS,\SeibergGMD} to a {\it complex} Wilson-Fisher scalar. So, assuming the validity of the $U(1)$ duality, the $SO$ duality cannot hold in this case, and thus also for other cases with $k=2$ and $N_f=N$.

For higher values of $k$ and $N=N_f=1$, the duality may be correct, namely the $SO(k)_{-\frac12}$ CS theory coupled to a single fermion may flow to the fixed point of a real Wilson-Fisher scalar. Again this implies that all baryonic operators of this theory decouple at low energies.

\subsec{The $k=2$ case}

In this case we have a duality between the theory of $N_f$ fermions coupled to $SO(2)_{-N+\frac{N_f}2}  =  U(1)_{-N+\frac{N_f}2}$, and the theory of $N_f$ real scalars coupled to $SO(N)_2$. Such a duality implies that the charge conjugation $\Z_2$ symmetry of the scalar theory is enhanced to $U(1)$, and its $SO(N_f)$ flavor symmetry is enhanced to $SU(N_f)$, at low energies.

As discussed above we can rule out the cases with $N_f = N$, so the lowest case is $N=2$ and $N_f=1$. This case is interesting because, as discussed around \SpecialFourDual, the two dual Abelian theories admit two more non-Abelian descriptions in which the full $SU(2)$ global symmetry of the fixed point is manifest in the UV.

Our duality maps Theory S, a $U(1)_2$ CS theory coupled to a complex Wilson-Fisher scalar (for $N_f=1$ there is no difference between the $SO(2)$ and $U(1)$ flows) to Theory F, a $U(1)_{-\frac32}$ CS theory coupled to a complex fermion. The very same two theories, viewed as $U(1)$ theories, are also mapped to each other by the $U(1)$ duality of \refs{\AharonyMJS,\SeibergGMD,\HsinBLU}. However, interestingly enough, the operator mappings are not the same in the $U(1)$ duality and the $SO(2)$ duality: the $U(1) \leftrightarrow U(1)$ duality preserves the magnetic symmetry and the charge conjugation, while the $SO(2) \leftrightarrow SO(2)$ duality exchanges them. Fortunately, this perfectly fits with the enhanced quantum $SU(2)$ global symmetry. In each $U(1)$ CS description there is a manifest $U(1)_J \rtimes \Z_2^\CC$ magnetic and charge conjugation symmetry. The $U(1)$ duality trivially maps the two copies of $U(1)_J \rtimes\Z_2^\CC$ one into the other. The $SO(2)$ duality, instead, maps $U(1)_J \rtimes\Z_2^\CC$ in a nontrivial way, which follows from its embedding inside the global $SU(2)$ symmetry: it is an $SU(2)$ rotation.

For $k=2$ and $N > 2$ we obtain more complicated dualities, which we cannot rule out. The $U(1)$ dualities map the fermionic $SO(2)  =  U(1)$ theories to $SU(N)_1$ theories coupled to $N_f$ scalars, so we obtain {\it boson-boson dualities} between $SU(N)_1$ and $SO(N)_2$ theories coupled to $N_f < N$ Wilson-Fisher scalars (which are complex and real, respectively).

\subsec{The $N=2$ case}

For $N=2$ we have an $SO(2)_k  =  U(1)_k$ CS theory coupled to $N_f$ charged scalars with Wilson-Fisher couplings; here we can have $N_f=1$ or $N_f=2$. The dual for $N_f=1$ is an $SO(k)_{-\frac32}$ theory coupled to a real fermion, and for $N_f=2$ an $SO(k)_{-1}$ theory coupled to two real fermions.
For $k=1$ and $k=2$ we have already discussed these theories above.

For $k > 2$ the dual theory is non-Abelian, and we cannot rule the duality out. Again, it implies that the charge conjugation symmetry of the fermionic theory should be enhanced to $U(1)$ at low energies, and its $SO(N_f)$ flavor symmetry to $SU(N_f)$.

The $U(1)$ duality with $N_f=1$ maps the same scalar theory to an $SU(k)_{-\frac12}$ theory coupled to a complex fermion, giving another {\it fermion-fermion duality} between $SO(k)_{-\frac32}$ and $SU(k)_{-\frac12}$ theories with one real/complex fermion flavor.
For $N_f=2$ the $U(1)$ duality breaks down, but the $SO(N)$ duality of the previous paragraph may still be valid.

\newsec{Relation to theories of high-spin gravity}

The $SO(N)$ theories with $k=0$ and $N_f=1$ were the first ones to be suggested to be dual at large $N$ to Vasiliev's high-spin gravity theory on $AdS_4$ \refs{\KlebanovJA,\SezginPT}; they are dual to the minimal Vasiliev theory, which has only even-spin excitations. There are two versions of this theory, differing by a discrete parameter, that were argued to be dual to theories of $N$ scalars and $N$ fermions, respectively. This was later generalized to $U(N)$ and $SU(N)$ theories being dual to non-minimal Vasiliev theories that have excitations of all spins. There is an obvious generalization of both dualities to higher $N_f$, with the $SO(N)$ theory containing $N_f(N_f+1)/2$ excitations of even spins, and $N_f(N_f-1)/2$ excitations of odd spins. The Vasiliev theory is only known by its classical equations of motion, so a priori it is not known how to quantize it; the dual field theories with finite $N$ can be viewed as giving a non-perturbative definition of this theory.

For the $U(N)$ scalar/fermion theories, it was argued that a parameter $\theta_0$ in the Vasiliev theory is related to $N/k$ when the $U(N)$ Yang-Mills theory is replaced by $U(N)_k$ \GiombiKC. This parameter interpolates between the theory dual to parity-invariant scalars, and the one related to parity-invariant fermions, and this led to the conjectured duality between CS-scalar and CS-fermion theories. The same parameter exists also in the minimal Vasiliev theories, so it is natural to conjecture that turning it on in the minimal Vasiliev theories corresponds to having $SO(N)_k$ or $USp(2N)_k$ CS-matter theories. Again the fact that the Vasiliev theory has two interpretations, as a CS-scalar and as a CS-fermion theory, suggests that at least at large $N$ the $SO(N)$ and $USp(2N)$ dualities that we discussed above are correct.

At leading order in large $N$, the orthogonal, symplectic and unitary theories are all the same, consistent with having the same classical equations of motion in the minimal and non-minimal Vasiliev theories (up to having a projection removing half of the fields in the $SO$ and $USp$ cases). However, the one-loop corrections should be different. In particular there should be a discrete parameter distinguishing the $SO(2N)$ and $USp(2N)$ theories, whose effect is to change the sign of all $l$-loop diagrams with odd values of $l$. This can be realized by inverting the signs of $\hbar$ and of Newton's constant.%
\foot{Note that this is not the same as the case where only Newton's constant is inverted, giving a theory in de Sitter space, as studied in \refs{\AnninosUI,\AnninosHIA}.}
There should also be discrete parameters on the gravity side distinguishing the different versions of the $SO(N)$ theories, where one gauges some of the $\Z_2$ discrete symmetries.

\newsec{Level-rank dualities with orthogonal and symplectic groups}

\subsec{Level-rank dualities of 3d TQFTs}

Level-rank dualities of 2d chiral algebras can be derived starting from systems of free fermions. For instance, consider a system of $Nk$ free real (Majorana) fermions: writing them as $\psi_{a\tilde a}$ with $a=1,\dots, N$ and $\tilde a=1,\dots, k$ one obtains the following conformal embeddings of chiral algebras (see also \refs{\Hasegawa,\VerstegenAT}):
\eqn\ConfEmbSpin{\eqalign{
Spin(Nk)_1 &\;\supset\; \big( Spin(N)_k \times Spin(k)_N \big)/\Z_2 \qquad N,k {\rm \ odd} \cr
Spin(Nk)_1 &\;\supset\; Spin(N)_k \times SO(k)_N \qquad\qquad\quad\; N {\rm \ even,\ } k {\rm \ odd} \cr
Spin(Nk)_1 &\;\supset\; SO(N)_k \times SO(k)_N \qquad\qquad\quad\;\;\;\, N,k {\rm \ even} \;.}}
Here $Spin$ is the standard Kac-Moody chiral algebra, while $SO  =  Spin/\Z_2$ is the extended chiral algebra \refs{\MSN,\MooreYH} obtained from $Spin$ by adding a suitable $\Z_2$ generator of spectral flow (see below). The $\Z_2$ quotient in the first line is the extension by the diagonal element. The centers match on the two sides. A series of equalities of chiral algebras follows:
\eqn\GKOSpin{\eqalign{
Spin(N)_k &\quad\longleftrightarrow\quad \frac{Spin(Nk)_1}{Spin(k)_N} \qquad\qquad N,k {\rm \ odd} \cr
Spin(N)_k &\quad\longleftrightarrow\quad \frac{Spin(Nk)_1}{SO(k)_N} \qquad\qquad N {\rm \ even,\ } k {\rm \ odd} \cr
SO(N)_k &\quad\longleftrightarrow\quad \frac{Spin(Nk)_1}{Spin(k)_N} \qquad\qquad N {\rm \ odd,\ } k {\rm \ even} \cr
SO(N)_k &\quad\longleftrightarrow\quad \frac{Spin(Nk)_1}{SO(k)_N} \qquad\qquad N,k {\rm \ even} \;.}}
On the right hand sides we have GKO cosets \GoddardVK. Moving from two-dimensional chiral algebras to three-dimensional Chern-Simons theories \MooreYH, one obtains dualities between the following Lagrangian theories:
\eqn\nonspinLRdualSpin{\eqalign{
Spin(N)_k &\quad\longleftrightarrow\quad \frac{Spin(Nk)_1 \times Spin(k)_{-N}}{\Z_2} \qquad\qquad N,k {\rm \ odd} \cr
Spin(N)_k &\quad\longleftrightarrow\quad Spin(Nk)_1 \times SO(k)_{-N} \qquad\qquad\quad N {\rm \ even,\ } k {\rm \ odd} \cr
SO(N)_k &\quad\longleftrightarrow\quad \frac{Spin(Nk)_1 \times Spin(k)_{-N}}B \qquad\qquad N {\rm \ odd,\ } k {\rm \ even} \cr
SO(N)_k &\quad\longleftrightarrow\quad \frac{Spin(Nk)_1 \times SO(k)_{-N}}{\Z_2} \qquad\qquad\;\;\, N,k {\rm \ even} \;.}}
On the right hand sides, the Lagrangian is the one corresponding to the Lie algebra of the numerator, while the gauge group is the result of the quotient. On the third line, $B = \Z_2 \times \Z_2$ for $k=0 \mod 4$ and $B=\Z_4$ for $k=2\mod 4$. We stress that these level-rank dualities of 3d TQFTs (or equivalently of 2d chiral algebras) can be rigorously proven. They have also been analyzed in \refs{\NaculichPA,\MlawerUV}.

Similarly, one can start with a system of $4Nk$ 2d real fermions, and writing them as $4Nk$ complex fermions with a symplectic Majorana condition one obtains the conformal embedding
\eqn\ConfEmbUSp{
Spin(4Nk)_1 \;\supset\; USp(2N)_k \times USp(2k)_N \;.}
This leads to the duality of chiral algebras
\eqn\GKOUSp{USp(2N)_k \quad\longleftrightarrow\quad \frac{Spin(4Nk)_1}{USp(2k)_N} \;,}
and in terms of three-dimensional Chern-Simons theories one has the duality
\eqn\nonspinLRdualUSp{
USp(2N)_k \quad\longleftrightarrow\quad \frac{Spin(4Nk)_1 \times USp(2k)_{-N}}{\Z_2} }
between 3d TQFTs.

\subsec{Matching the symmetries}

Before going on with the analysis of those dualities, let us fix some notations for orthogonal and symplectic chiral algebras. The center $B(G)$ of the simply-connected group $G$ associated to a Lie algebra $\frak g$ acts on the affine Lie algebra $G_k$ as an outer automorphism, and its action is generated by elements $\sigma_i$ of $G_k$ via spectral flow. In the corresponding 3d CS theory, $B(G)$ appears as a one-form symmetry \GaiottoKFA\ generated by the lines $\sigma_i$. Such lines can act in two different ways on the other lines of the theory: either by ``fusion'' (or spectral flow), if they are placed parallel to the lines they act upon, or by ``monodromy'', if they are wound on a small circle around the other lines.

Whenever the generator of spectral flow has integer dimension, it can be included in the chiral algebra to give an extended chiral algebra; equivalently, in 3d the one-form symmetry can be gauged to give a $G_k/H$ CS TQFT, where $H$ is a subgroup of $B(G)$.
When the generator has half-integer dimension, the chiral algebra can be augmented to a $\Z_2$-graded chiral algebra (with half-integer dimensions) that depends on the spin structure; equivalently, in 3d the one-form symmetry can be gauged but the $G_k/H$ CS theory is a spin-TQFT. In our case, the center of $USp(2N)$ is $\Z_2$ while the center of $Spin(N)$ is $\Z_2$ for $N$ odd, $\Z_2 \times \Z_2$ for $N=0\mod4$ and $\Z_4$ for $N=2\mod 4$. Let us identify the corresponding generators of spectral flow.

In $USp(2N)_k$, the $\Z_2$ spectral flow is generated by%
\foot{We indicate a highest weight representation by its Dynkin labels $(\lambda_1, \dots, \lambda_r)$ or by its extended Dynkin labels $[\lambda_0, \dots, \lambda_r]$, where $r$ is the rank. In $\frak s \frak p(2N)$, $\lambda_r$ refers to the long root. In $\frak s \frak o(2n+1)$, $\lambda_r$ refers to the short root, while in $\frak s \frak o(2n)$, $\lambda_{r-1}$ and $\lambda_r$ refer to the two roots at the ``bifurcated tail'' of the Dynkin diagram.}
\eqn\DefSigmaUSp{
\sigma: [\lambda_0, \lambda_1, \dots, \lambda_N] \to [\lambda_N, \lambda_{N-1}, \dots, \lambda_0] \;. }
The generator is given by $\sigma = (0, \dots, 0 , k)$ with dimension $h(\sigma) =  \frac{kN}4$. The action of $\sigma$ via monodromy is
\eqn\MonSigmaUSp{
Q_\sigma [\lambda] = (-1)^c \, [\lambda] \;, }
where the congruence class $c$ of a representation $[\lambda]$ is given by $c = \sum_{j {\rm \ odd}}^N \lambda_j \mod 2$. In particular the self-parity of $\sigma$ is $(-1)^{Nk}$. For $Nk=0\mod 4$, one can consider the $PUSp(2N)_k \equiv USp(2N)_k/\Z_2$ CS theory; for $Nk=2\mod 4$, one can consider the $PUSp(2N)_k$ spin-CS theory.

In $Spin(N)_k$ with $N$ odd, the $\Z_2$ spectral flow is generated by
\eqn\DefSigmaSpinOdd{
\sigma: [\lambda_0, \lambda_1, \lambda_2 \dots, \lambda_r] \to [\lambda_1, \lambda_0, \lambda_2, \dots, \lambda_r] \;. }
The generator is given by $\sigma = (k,0, \dots, 0)$ with dimension $h(\sigma) = \frac k2$. The action via monodromy is
\eqn\MonSigmaSpinOdd{
Q_\sigma \, [\lambda] = (-1)^{\lambda_r} \, [\lambda] \;. }
For $N$ even we have the two spectral flow operations
\eqn\DefJsSpinEven{
j_s:[\lambda_0, \lambda_1, \dots, \lambda_{r-1}, \lambda_r] \to \cases{ [\lambda_{r-1}, \lambda_r, \lambda_{r-2}, \dots, \lambda_1, \lambda_0] &for $N=2\mod 4$ \cr [\lambda_r, \lambda_{r-1}, \lambda_{r-2}, \dots, \lambda_1, \lambda_0] &for $N=0 \mod 4$ } }
and
\eqn\DefSgiamSpinEven{
\sigma: [\lambda_0, \lambda_1, \dots, \lambda_{r-1}, \lambda_r] \to [\lambda_1, \lambda_0, \lambda_2, \dots, \lambda_{r-2} , \lambda_r, \lambda_{r-1}] \;. }
They are generated by $j_s = (0, \dots, 0,k)$ with dimension $h(j_s) = \frac{Nk}{16}$ and $\sigma = (k, 0, \dots, 0)$ with dimension $h(\sigma) = \frac k2$. For $N=2 \mod 4$ the group structure is $\Z_4$: $j_s^2 = \sigma$ and $j_s^4 = \sigma^2 = \unit$. The action via monodromy is
\eqn\MonJsSpinEvenI{
Q_{j_s}[\lambda] = i^{\frac N2 c} \, [\lambda] }
where the congruence class is $c = \sum_{j {\rm \ odd}}^{\frac N2-2}2 \lambda_j + \lambda_r - \lambda_{r-1} \mod4$. For $N=0\mod 4$ the group structure is $\Z_2 \times \Z_2$: $j_s^2 = \sigma^2 =\unit$ and $j_s \sigma = \sigma j_s$. The action via monodromy is
\eqn\MonJsSpinEvenII{
Q_{j_s} [\lambda] = \cases{ (-1)^{c_c} [\lambda] &for $N = 0 \mod 8$ \cr (-1)^{c_s} [\lambda] &for $N=4 \mod 8$ } }
where $c_s = \sum_{j {\rm \ odd}}^{\frac N2-3} \lambda_j + \lambda_r \mod 2$ and $c_c = \sum_{j{\rm \ odd}}^{\frac N2-3} + \lambda_{r-1} \mod 2$. The action of $\sigma$ via monodromy is
\eqn\MonSigmaSpinEven{
Q_\sigma [\lambda] = (-1)^{\lambda_r + \lambda_{r-1}} \, [\lambda] }
in both $N$ even cases. The one-form symmetry generated by $\sigma$ can always be gauged to obtain the $SO(N)_k$ CS theory: it is a TQFT for $k$ even, and a spin-TQFT for $k$ odd. Only for $N,k$ both even there is another generator $j_s$ that survives the quotient, thus only in this case $SO(N)_k$ has a $\Z_2$ one-form symmetry.

Let us also discuss what type of conventional zero-form symmetries the Chern-Simons theories can have. In $Spin(N)_k$ and $SO(N)_k$ with $N$ even, one defines a ``charge conjugation'' $\Z_2$ symmetry $\CC$ that transforms representations as
\eqn\DefC{
\CC:\; \lambda_{r-1} \;\leftrightarrow\; \lambda_r \;. }
In $SO(N)_k$ gauging this symmetry gives $O(N)_k$. Counterterms for the classical gauge field of $\CC$ lead to additional parameters in the $O(N)_k$ theory \AharonyKMA. In $SO(N)_k$ with $k$ even, we define a ``magnetic'' $\Z_2$ symmetry $\CM$ that exchanges the two representations of the extended chiral algebra resulting from a fixed point of the $\Z_2$ spectral flow of $Spin(N)_k$ \refs{\MSN,\MooreYH}. From the 3d point of view, the magnetic quantum number of a monopole operator is the second Stiefel-Whitney class $w_2$ of the $SO(N)$ bundle around its location. This symmetry is gauged when going from $SO(N)_k$ to $Spin(N)_k$.

Whenever a three-dimensional TQFT has a $\Z_2$ one-form global symmetry generated by $\sigma$ with spin $h(\sigma) = \frac12 \mod 1$, we can define a quantum zero-form symmetry $\CK_\sigma$ acting on the lines in the following way:
\eqn\DefK{
\CK_\sigma [\lambda] = \cases{ [\lambda] &if $Q_\sigma [\lambda] = [\lambda]$ \cr \sigma \cdot [\lambda] &if $Q_\sigma [\lambda] = - [\lambda]$ , } }
where by $\sigma\cdot [\lambda]$ we mean fusion. This definition guarantees that $\CK$ preserves the fusion rules and that $\CK[\lambda]$ has the same spin as $[\lambda]$.

Having settled the basic definitions, we can analyze the precise mapping of symmetries between the dual theories in \nonspinLRdualSpin\ and \nonspinLRdualUSp.

Consider
\eqn\SpinDualNoKo{
Spin(N)_k \quad\longleftrightarrow\quad \frac{ Spin(Nk)_1 \times Spin(k)_{-N} }{\Z_2} \qquad\qquad N,k {\rm \ odd} \;, }
where the quotient is generated by $\sigma \otimes \sigma$ (which has fixed points). Both sides have a $\Z_2$ one-form global symmetry, and the map of generators is%
\foot{Here and in the following, $\sim$ means identification by the quotient.}
$\sigma \leftrightarrow \sigma \otimes \unit \sim \unit \otimes \sigma$. Both sides have $\Z_2$ zero-form symmetry. On the RHS it is the quantum symmetry $\CK_\sigma$. On the LHS it is the magnetic symmetry $\CM_{\Z_2}$ associated to the fixed points of the $\Z_2$ quotient. The map of generators is $\CK_\sigma \leftrightarrow \CM_{\Z_2}$.

Consider
\eqn\SpinDualNeKo{
Spin(N)_k \quad\longleftrightarrow\quad Spin(Nk)_1 \times SO(k)_{-N} \qquad\qquad N {\rm \ even,\ } k {\rm \ odd} \;. }
Both sides have a $\Z_2 \times \Z_2$ one-form global symmetry for $N=0 \mod 4$, and $\Z_4$ for $N=2\mod 4$. The map of generators is $j_s \leftrightarrow j_s \otimes \unit$, $\sigma \leftrightarrow \sigma\otimes \unit$. Both sides have a $\Z_2 \times \Z_2$ zero-form symmetry, and the map of generators is $\CC\CK_\sigma \leftrightarrow \unit \otimes \CM$, $\CK_\sigma \leftrightarrow \CC \otimes \unit$.

Consider
\eqn\SpinDualNoKe{
SO(N)_k \quad\longleftrightarrow\quad \frac{Spin(Nk)_1 \times Spin(k)_{-N}}{B} \qquad\qquad N {\rm \ odd,\ } k {\rm \ even} \;. }
For $k=2 \mod 4$, $B=\Z_4$ is generated by $j_s \otimes j_s$, while for $k = 0 \mod 4$, $B = \Z_2 \times \Z_2$ is generated by $j_s \otimes j_s$ and $\sigma \otimes \sigma$ (the quotient has no fixed points). Both sides have no zero-form symmetry (on the RHS, all generators of the numerator are projected out by the quotient). The zero-form symmetry is $\Z_2$ and the map of generators is $\CM \leftrightarrow \CC\CK_\sigma$ (on the RHS, all other generators do not commute with the quotient action and are thus broken).

Consider
\eqn\SpinDualNeKe{
SO(N)_k \quad\longleftrightarrow\quad \frac{Spin(Nk)_1 \times SO(k)_{-N}}{\Z_2} \qquad\qquad N,k {\rm \ even} \;, }
where the quotient is generated by $j_s \otimes j_s$ (with no fixed points). On both sides the one-form global symmetry is $\Z_2$ and the map of generators is
\eqn\SpinDualNeKeOpMap{
j_s \;\leftrightarrow\; \cases{ j_s \otimes \unit \;\sim\; \unit \otimes j_s \qquad& $\frac{Nk}4$ even \cr
\sigma j_s \otimes \unit \;\sim\; \sigma \otimes j_s \qquad& $\frac{Nk}4$ odd. }}
On the RHS, all other generators are projected out by the quotient. The zero-form symmetry is $\Z_2 \times \Z_2$, and the map of generators is $\CC \leftrightarrow \unit \otimes \CM$, $\CM \leftrightarrow \unit\otimes \CC$.

Finally, consider
\eqn\USpDual{
USp(2N)_k \quad\longleftrightarrow\quad \frac{Spin(4Nk)_1 \times USp(2k)_{-N}}{\Z_2} \;, }
where the quotient is generated by $j_s \otimes \sigma$ (with no fixed points). On both sides the one-form symmetry is $\Z_2$ and the map of generators is
\eqn\USpDualOpMap{
\sigma \;\leftrightarrow\; \cases{ j_s \otimes \unit \;\sim\; \unit \otimes \sigma \qquad& $Nk$ even \cr
\sigma j_s \otimes \unit \;\sim\; \sigma \otimes \sigma \qquad& $Nk$ odd. }}
The zero-form symmetry is $\Z_2$ for $Nk=2\mod 4$ and nothing otherwise. The map of generators is $\CK_\sigma \leftrightarrow \unit \otimes \CK_\sigma$.

\subsec{Level-rank dualities of spin-TQFTs}

So far we have discussed level-rank dualities between TQFTs. We can obtain simpler dualities if we consider spin-TQFTs. As we explain below, we obtain the following:
\eqnn\spinLRdualSO
\eqnn\spinLRdualUSp
$$\eqalignno{
SO(N)_k \times SO(0)_1 &\quad\longleftrightarrow\quad SO(k)_{-N} \times SO(Nk)_1 &\spinLRdualSO \cr
USp(2N)_k \times SO(0)_1 &\quad\longleftrightarrow\quad USp(2k)_{-N} \times SO(4Nk)_1 \;. &\spinLRdualUSp
}$$
We recall that $SO(N)_1$ is a trivial spin-TQFT with two transparent lines of spins $\big\{0, \frac12\big\}$ and with framing anomaly $c = \frac{N}2$ (see e.g. \SeibergRSG).

Before deriving \spinLRdualSO\ and \spinLRdualUSp, let us discuss the symmetries and their map, starting with the orthogonal case \spinLRdualSO. As we discussed after \MonSigmaSpinEven, $SO(N)_k$ has a $\Z_2$ one-form global symmetry for $N,k$ both even, and not otherwise. Thus the one-form symmetries match. Moreover, $SO(N)_k$ has a charge conjugation $\Z_2$ zero-form symmetry $\CC$ for $N$ even, and a magnetic $\Z_2$ symmetry $\CM$ for $k$ even.%
\foot{For $k$ even, the $\Z_2$ quotient $Spin(N)_k/\Z_2  =  SO(N)_k$ has fixed points, but not for $k$ odd when $SO(N)_k$ is spin.}
Those two symmetries are exchanged in the duality \spinLRdualSO,
\eqn\CMexchange{
\CC \quad\longleftrightarrow\quad \CM \;, }
as it also follows from the derivation of the duality that we give below.

In the symplectic case \spinLRdualUSp, on both sides there is a $\Z_2$ one-form global symmetry generated by $\sigma$, and a quantum zero-form symmetry $\CK_\sigma$ for $Nk = 2\mod 4$.

Note that in the CS theories with matter in the fundamental representation discussed in the main text, both for gauge group $SO$ and $USp$, the possible one-form global symmetry is broken by the presence of matter \GaiottoKFA.

Next, we derive the dualities \spinLRdualSO\ and \spinLRdualUSp. The simplest case is to start with $Spin(N)_k \leftrightarrow Spin(Nk)_1 \times SO(k)_{-N}$ with $N$ even, $k$ odd. We can gauge the $\Z_2$ one-form symmetry generated by $\sigma \leftrightarrow \sigma\otimes\unit$ to directly obtain \spinLRdualSO. Since $k$ is odd, $SO(N)_k$ is a spin theory and hence adding $SO(0)_1$ does not change it. Another simple case is to start with $Spin(N)_k \leftrightarrow \big( Spin(Nk)_1 \times Spin(k)_{-N} \big)/\Z_2$ with $N,k$ odd. The quotient is by $\sigma \otimes \sigma$, and it preserves the $\Z_2$ generator $\sigma \leftrightarrow \sigma\otimes \unit \sim \unit \otimes \sigma$. If we gauge the latter one-form symmetry as well, we obtain
\eqn\unnsospa{
SO(N)_k \quad\longleftrightarrow\quad \frac{Spin(Nk)_1}{\Z_2} \times \frac{Spin(k)_{-N}}{\Z_2} \;,
}
which is precisely \spinLRdualSO.

To get the other cases, we make use of the following identity of spin-TQFTs:
\eqn\ZtwoUnfoldingRel{
Spin(2L)_1 \times SO(0)_1 \quad\longleftrightarrow\quad SO(2L)_1 \times (\Z_2)_{-L} \;, }
discussed in \SeibergRSG. Starting from $USp(2N)_k \leftrightarrow \big( Spin(4Nk)_1 \times USp(2k)_{-N} \big)/\Z_2$, we multiply both sides by $SO(0)_1$ making them into spin theories, then apply \ZtwoUnfoldingRel\ and obtain
\eqn\unnumbUSPSOa{
USp(2N)_k \times SO(0)_1 \quad\longleftrightarrow\quad \frac{SO(4Nk)_1 \times (\Z_2)_{-2Nk} \times USp(2k)_{-N}}{\Z_2} \;.
}
The theory $(\Z_2)_{-2Nk}$ is a TQFT with $\Z_2 \times \Z_2$ one-form symmetry. The $\Z_2$ quotient is generated by pairing $\sigma$ in $USp(2k)_{-N}$, whose spin is $h = - \frac{Nk}4 \mod 1$, with a generator in $(\Z_2)_{-2Nk}$ that has opposite spin. In the quotient $SO(4Nk)_1$ remains as a spectator. The product of $USp(2k)_{-N}$ by $(\Z_2)_{-2Nk}$ gives four times as many fields, however the freely-acting quotient by $\Z_2$ reduces to the original ones (one can check that the surviving states have the same dimensions as the original ones). Thus one has a simple duality of TQFTs: $\big( (\Z_2)_{-2Nk} \times USp(2k)_{-N} \big)/\Z_2 \leftrightarrow USp(2k)_{-N}$. This leads to the duality in \spinLRdualUSp. Exactly the same reasoning can be applied to the case $SO(N)_k \leftrightarrow \big( Spin(Nk)_1 \times SO(k)_{-N} \big)/\Z_2$ with $N,k$ even. We multiply both sides by $SO(0)_1$, then we apply \ZtwoUnfoldingRel, and finally observe that the quotient can be ``unfolded'' by the simple duality: $\big( (\Z_2)_{-2nm} \times SO(2m)_{-2n} \big)/\Z_2 \leftrightarrow SO(2m)_{-2n}$. Here we set $N = 2n$ and $k=2m$. This leads to \spinLRdualSO.

The last case is $SO(N)_k \leftrightarrow \big( Spin(Nk)_1 \times Spin(k)_{-N} \big)/B$, with $N$ odd, $k$ even. After multiplication by $SO(0)_1$ and application of \ZtwoUnfoldingRel, we obtain
\eqn\unnSOSOSO{
SO(N)_k \times SO(0)_1 \quad\longleftrightarrow\quad SO(Nk)_1 \times \frac{(\Z_2)_{-\frac{Nk}2} \times Spin(k)_{-N}}B \;.
}
This case is a little bit more intricate. For $Nk=0 \mod 4$, $B = \Z_2^{(j_s)} \times \Z_2^{(\sigma)}$. The first factor is obtained by pairing $j_s$ in $Spin(k)_{-N}$, whose spin is $h(j_s) = -\frac{Nk}{16} \mod 1$, with a generator in $(\Z_2)_{-\frac{Nk}2}$ that has the opposite spin. Therefore the first quotient is non-spin. The second factor is generated by $W_{1,0} \otimes \sigma$ (in the notation of \SeibergRSG) and it is a spin quotient. We can first use $(\Z_2)_{-\frac{Nk}2}$ to unfold the quotient by $\Z_2^{(j_s)}$, and be left with $Spin(k)_{-N} / \Z_2^{(\sigma)} = SO(k)_{-N}$. This reproduces \spinLRdualSO. For $Nk=2\mod 4$, $B = \Z_4$. Its generator is obtained by pairing $j_s$ in $Spin(k)_{-N}$ with a generator in $(\Z_2)_{-\frac{Nk}2}$ that has the opposite spin. This quotient has no fixed points, and its effect is to restrict to the lines with $c=0\mod 4$ with respect to $j_s$, times the transparent lines $\big\{0, \frac12\big\}$. This is precisely the effect of the spin quotient $Spin(k)_{-N} / \Z_2^{(\sigma)}$ for $N$ odd. Again we end up with \spinLRdualSO.

\subsec{More non-spin level-rank dualities}

For special values of $N,k$ the spin dualities \spinLRdualSO-\spinLRdualUSp\ can be upgraded to non-spin dualities.
This uses the fact when $N=0$ mod 8, the spin duality \ZtwoUnfoldingRel\ can be replaced by the non-spin duality
\eqn\spinsixteen{
Spin(16L)_1 \quad\longleftrightarrow\quad (\Z_2)_0\times (E_8)^L_1 \;.
}
We will omit the trivial TQFT $(E_8)_1$.
For even $N,k$ and $Nk=0$ mod 16, repeating the derivation of \spinLRdualSO\ with \spinsixteen\ replacing \ZtwoUnfoldingRel\ gives the non-spin duality
\eqn\sononspin{
SO(N)_k \quad\longleftrightarrow\quad SO(k)_{-N} \;.
}
Similarly, for $Nk=0$ mod 4, repeating the derivation of \spinLRdualUSp\ with \spinsixteen\ gives the non-spin duality
\eqn\spnonspin{
USp(2N)_k \quad\longleftrightarrow\quad USp(2k)_{-N} \;.
}

A special case of \sononspin\ is $SO(8L)_2 \leftrightarrow SO(2)_{-8L}$, which can be rewritten as $SU(8L)_1 \leftrightarrow U(1)_{-8L}$ using some relation in Appendix A. In turn this can be used in the arguments of \HsinBLU\ to show the non-spin duality
\eqn\sunonspin{
SU(N)_k \quad\longleftrightarrow\quad U(k)_{-N}
}
for even $N$ and $Nk = 0 \mod 8$.

In the next section we will make use of the non-spin level-rank dualities \sononspin\ and \spnonspin\ to find new $\CT$-invariant TQFTs.

\newsec{${\cal T}$-invariant TQFTs from level-rank duality}

It is of considerable interest to find topological field theories that are time-reversal invariant ($\CT$-invariant) up to an anomaly, because they can lead to gapped boundary states of topological insulators or topological superconductors.  Some known examples are based on Chern-Simons theories with various product groups and possibly appropriate quotients \refs{\FidkowskiJUA\WangUKY\ChenJHA\BondersonPLA\MetlitskiBPA\WangQMT-\MetlitskiEKA,
\SeibergRSG}. It turns out that level-rank duality is a powerful tool to find new examples.\foot{We thank E.~Witten for a useful discussion about this point.} Specifically, a level-rank duality that exchanges $N \leftrightarrow k$, when applied to a theory with $N=k$, shows that such a theory is $\CT$-invariant quantum mechanically.

Let us examine a few examples. First, from the dualities of spin-TQFTs (4.15), (4.19) in \HsinBLU, and \spinLRdualSO, \spinLRdualUSp\ here, we find spin-TQFTs that are $\CT$-invariant, up to an anomaly. In some cases the theory involved is already a spin theory. In some other cases the theory is not spin but the level-rank duality, and therefore $\CT$-invariance, only holds after we tensor with a trivial spin theory, which we denote by $\psi$. Second, from the dualities of non-spin TQFTs (4.18) in \HsinBLU\ and \sononspin, \spnonspin\ here we find conventional TQFTs that are $\CT$-invariant, up to an anomaly. This leads to the following examples:

\bigskip

\halign{
\hfil#\hfil&\hfil#\hfil&
\hfil#\hfil&
\hfil#\hfil&
\hfil#\hfil \cr
\noalign{\hrule\smallskip}
\omit\bf $\CT$-invariant theories \kern-5pt &  ${\bf N}$ & {\bf property} & {\bf framing anomaly} \cr
\noalign{\smallskip\hrule\smallskip}
$U(N)_{N,2N}$ & even $N$ & ${\cal T}$-invariant spin theory & $(N^2+1)/2$ \cr
 & odd $N$ & need to add $\psi$ & \cr
\noalign{\smallskip\hrule\smallskip}
$PSU(N)_N$ & even $N$ & ${\cal T}$-invariant spin theory & $(N^2-1)/2$ \cr
 & odd $N$ & ${\cal T}$-invariant non-spin theory & \cr
\noalign{\smallskip\hrule\smallskip}
$USp(2N)_N$ & even $N$ & ${\cal T}$-invariant non-spin theory & $N^2$ \cr
	& odd $N$ & need to add $\psi$ &  \cr
\noalign{\smallskip\hrule\smallskip}
$SO(N)_N$ & odd $N$ & ${\cal T}$-invariant spin theory & \cr
& $N=0$ mod 4 \kern10pt & ${\cal T}$-invariant non-spin theory & $N^2/4$\cr
& $N=2$ mod 4 \kern10pt & need to add $\psi$ & \cr
\noalign{\smallskip\hrule}
}

\bigskip

The special case $SO(3)_3$ appears in the literature as $SU(2)_6/\Z_2$ \FidkowskiJUA.
The special case $U(1)_2$ is known as the ``fermion/semion theory'' \FidkowskiJUA\ and was discussed recently in \refs{\SeibergRSG,\SeibergGMD}.

\bigskip
\noindent{\bf Acknowledgments}

We would like to thank E.~Witten and R.~Yacoby for useful discussions.
The work of OA was supported in part  by the I-CORE program of the Planning and Budgeting Committee and the Israel Science Foundation (grant number 1937/12), by an Israel Science Foundation center for excellence grant, by the Minerva foundation with funding from the Federal German Ministry for Education and Research, by a Henri Gutwirth award from the Henri Gutwirth Fund for the Promotion of Research, and by the ISF within the ISF-UGC joint research program framework (grant no.\ 1200/14). OA is the Samuel Sebba Professorial Chair of Pure and Applied Physics.
FB is supported in part by the MIUR-SIR grant RBSI1471GJ ``Quantum Field Theories at Strong Coupling: Exact Computations and Applications'', by the INFN, and by the IBM Einstein Fellowship at the Institute for Advanced Study.
The work of PH is supported by Physics Department of Princeton University.
NS was supported in part by DOE grant DE-SC0009988.


\appendix{A}{Notations and useful facts about Chern-Simons theories}

Let us start by reviewing some facts about the fermion determinant \refs{\AlvarezGaumeNF,\WittenABA} and our notation.
A $(2+1)d$ fermion coupled to a gauge field $A$ and transforming in a complex or pseudo-real representation ${\bf r}$ has partition function
\eqn\fermionZ{
Z_\psi = |Z_\psi | \, \exp\big(-i \pi \eta(A)/2\big) \;,
}
where $\eta(A)$ is the eta-invariant of the Dirac operator and the sign in the exponent is a matter of convention. For a fermion in a real representation, the phase of the partition function is $\exp\big( -i\pi \eta(A)/4 \big)$ instead.

By the Atiyah-Patodi-Singer index theorem \AtiyahJF,
\eqn\APS{
e^{i\pi\eta(A)}
=\exp\left(2\pi i\int_X \hat A(R) \Tr\nolimits_{{\bf r}} e^{F/ 2\pi}\right)
=\exp\left( 2ix_{{\bf r}} \int {\rm CS}(A) + 2i({\rm dim }\ {\bf r})\int {\rm CS_{grav}}
\right)
}
where $X$ is a bulk four-manifold that bounds the three-manifold, $x_{{\bf r}}$ is the Dynkin index of the representation ${\bf r}$,%
\foot{For $SU(N)$, the fundamental has $x_{{\bf r}}=\half$ and the adjoint has $x_{{\bf r}}=N$.  For $SO(N)$, the vector has $x_{{\bf r}}=1$ and the adjoint has $x_{{\bf r}}=N-2$. For $USp(2N)$, the fundamental has $x_{{\bf r}} = \half$ and the adjoint has $x_{{\bf r}}=N+1$.}
and $\int {\rm CS_{grav}}=\pi\int_X \hat A(R)$ is the gravitational Chern-Simons term. In particular $e^{-in\int {\rm CS_{grav}}}$ is the partition function of the almost trivial spin-TQFT $SO(n)_1$.

The Lagrangian of a Chern-Simons-matter theory can include a bare Chern-Simons term $k_{\rm bare}$. This must be properly normalized. For example, if the gauge group is $SU(N)$ we have $k_{\rm bare}\in \Z$, while in $SU(N)/\Z_N$ we must have $k_{\rm bare}\in N\Z$.  In \Oferdg, \Oferdgo\ and below we define the level $k$ as the sum of the bare value and the possibly fractional value $\delta k$ that comes from \fermionZ:
\eqn\level{
k=k_{\rm bare}-\delta k \;,
}
where $\delta k=x_{{\bf r}}$ for complex or pseudo-real representations, and $\delta k={1\over 2}x_{{\bf r}}$ for real representations.

Let us collect some useful formulas. First, in our notation:
\eqn\CSfacts{\eqalign{
SO(2)_k &=U(1)_k \;,\qquad\qquad\qquad Spin(2)_k=U(1)_{4k} \cr
SO(3)_k &=SU(2)_{2k}/\Z_2 \;,\qquad\quad\; Spin(3)_k=SU(2)_{2k} \;.
}}
Second, chiral algebras at small level satisfy special relations:
\eqn\CSlowlevel{\eqalign{
SO(N)_2 &\quad\leftrightarrow\quad SU(N)_1 \;.
}}
Finally, the theories of a $U(1)$ or $\Z_2$ gauge field with an action expressed in terms of the eta-invariant are dual to simple spin-TQFTs:
\eqn\CSeta{\eqalign{
SU(N)_1\times SO(0)_1 &\quad\leftrightarrow\quad S = - N\pi \, \eta\big( U(1)\ {\rm gauge\ field}\big) \cr
Spin(N)_1\times SO(0)_1 &\quad\leftrightarrow\quad S = -{N\pi\over 2}\, \eta\big( \Z_2\ {\rm gauge\ field}\big) \;,
}}
where $SO(1)_1$ and $Spin(1)_1$ are the spin and non-spin Ising TQFT, respectively \SeibergRSG.
These dualities are proven in \HsinBLU\ and \SeibergRSG.


\listrefs
\end